\documentclass[10pt,final]{IEEEtran}
\usepackage{amsfonts}
\usepackage{mathrsfs}
\usepackage{amssymb}
\usepackage{graphicx}
\usepackage{psfrag}
\usepackage{amsmath}
\usepackage{array}
\usepackage{cases}
\usepackage{cite}
\usepackage{tabularx}
\usepackage{url}
\usepackage{color}
\usepackage{soul}
\usepackage[level=0]{wgroup_message}
\definecolor{mblue}{rgb}{0.05,0.05,0.6}
\makeatletter
\def\hlinew#1{%
  \noalign{\ifnum0=`}\fi\hrule \@height #1 \futurelet
   \reserved@a\@xhline}
\makeatother

\newcommand{\mc}[2]{{#1}^{(\!\!\;#2\!\!\;)}}
\newcommand{\mnc}[3]{{#1}^{(\!\!\;#2\!\!\;)}_{#3}}
\newcommand{\MNC}[3]{{\mathbf{#1}}^{(\!\!\;#2\!\!\;)}_{#3}}

\newcommand{\savespacess}{\vspace{-1.5pt}}
\newcommand{\skiph}{\hspace{-2mm}}

\newcommand{\ml}{\hspace{-.02mm}\ell\hspace{0.3mm}}

\newcommand{\dis}{\displaystyle}
\makeatletter
\renewcommand*{\@opargbegintheorem}[3]{\@IEEEtmpitemindent\itemindent\topsep 0pt\rmfamily \trivlist%
      \item[\hskip \labelsep{\indent\itshape #1\ #2}] \textit{(#3):}\ \itemindent\@IEEEtmpitemindent}
\makeatother

\newcommand{\paperTitle}{Generalized Interference Alignment --- Part II: \\Application to Wireless Secrecy}

\usepackage{varioref}

\begin{document}

\author{Liangzhong~Ruan,~\IEEEmembership{Member,~IEEE},
Vincent~K.N.~Lau, \IEEEmembership{Fellow,~IEEE},
and Moe~Z.~Win, \IEEEmembership{Fellow,~IEEE}

\thanks{L.\ Ruan and M.\ Z.\ Win are with the Laboratory for Information and Decision Systems
(LIDS), MIT (e-mail: \texttt{lruan, moewin@mit.edu)}.}
\thanks{V.\ K.\ N.\  Lau is with the ECE Department, HKUST (e-mail: \texttt{eeknlau@ust.hk}).}
}

\title{\paperTitle}

\newtheorem{Thm}{Theorem}[section]
\newtheorem{Thm2}{Theorem}[subsection]
\newtheorem{Lem}{Lemma}[section]
\newtheorem{Lem2}{Lemma}[subsection]
\newtheorem{Asm}{Assumption}[section]
\newtheorem{Def}{Definition}
\newtheorem{Remark}{Remark}[section]
\newtheorem{Prob}{Problem}[section]
\newtheorem{Prop}{Proposition}[section]
\newtheorem{Prop2}{Proposition}[subsection]
\newtheorem{Alg}{Algorithm}
\newtheorem{Scen}{Scenario}
\newtheorem{Exa}{Example}
\newtheorem{Cor}{Corollary}[section]
\definecolor{mblue}{rgb}{0.05,0.05,0.6}

\maketitle
\begin{abstract}
In contrast to its wired counterpart, wireless communication is highly susceptible to eavesdropping due to the broadcast nature of the wireless propagation medium.
Recent works have proposed  the use of interference to reduce eavesdropping capabilities in wireless wiretap networks.
However, the concurrent effect of interference on both eavesdropping receivers (ERs) and  legitimate receivers (LRs) has not been thoroughly investigated, and carefully engineering the network interference is required to harness the full potential of interference for wireless secrecy.
This two part paper addresses this issue by proposing a generalized interference alignment (GIA) technique, which jointly designs the transceivers at the legitimate partners to impede the ERs without interfering with LRs.
In Part I, we have established a theoretical framework for the GIA technique.
In Part II, we will first propose an efficient GIA algorithm that is applicable to large-scale networks and then evaluate the performance of this algorithm in stochastic wireless wiretap network via both analysis and simulation. These results reveal insights into when and how GIA contributes to wireless secrecy.
\end{abstract}
\renewcommand{\IEEEQED}{\IEEEQEDopen}

\section{Introduction}
\subsection{Background and Survey}
The ability to exchange confidential messages securely in wireless network has become increasingly important for modern information society.
However, in contrast to its wired counterpart, wireless transmission is highly susceptible to eavesdropping due to the broadcast nature of the wireless propagation medium\cite{BloBarRodMcl:08}.
Contemporary wireless security systems, based on cryptographic primitives, evolved from schemes developed for traditional wireline applications.
To overcome challenges associated with broadcast communication, one must augment contemporary wireless security techniques using strategies that exploit the intrinsic properties of the wireless propagation medium.
\mysubnote{Interference is a promising tool promote secrecy,}

\mysubnote{yet existing works apply to small networks only.}

A key observation in exploiting these properties is that the broadcast nature generates contrasting effects:
It makes the secrecy information from a certain legitimate transmitter (LT) vulnerable to malicious interception,
but at the same time enables other legitimate partners to impede the ERs via \emph{interference}.
Therefore, interference emerges as a potentially valuable resource for wireless network secrecy \cite{RabConWin:J15}.
The idea of enhancing network secrecy through the use of interference has been investigated in several recent works, under the name of artificial noise\cite{GoeNeg:08,GoeVasTowAdaDinLeu:11}, artificial noise alignment\cite{FakJafSwi:11,KhiZha:13}, friendly jamming \cite{SonHanJiaDeb:10,VilBloBarMcl:11}, or cooperative jamming  \cite{TekYen:08,DonHanPetPoo:10,HuaSwi:11,LiPetWeb:11}. A major challenge in utilizing interference to enhance secrecy is that while impeding the ERs, interference affects the LRs as well.
Hence, without proper coordination, interference may be of little help or even harmful to wireless secrecy in some network configurations \cite{VilBloBarMcl:11}.
We envision that a greater secrecy gain will be achieved by simultaneously coordinating
multiple legitimate partners such that aggregated interference causes negligible effects at the LRs while impeding ERs.
This motivates the need to develop coordinative interference engineering strategies for wireless wiretap networks, which will be referred to as wireless-tap networks.\footnote{ ``wireless wiretap" is referred to as ``wireless-tap" to emphasize the wireless nature of the propagation medium.}
\mysubnote{interference alignment is a good candidate, but prior works are based on an impractical approach.}

Several secrecy-enhancing interference engineering strategies have been proposed  for small networks with one LT\cite{FakJafSwi:11,KhiZha:13,VilBloBarMcl:11,TekYen:08} or one LR \cite{SonHanJiaDeb:10,DonHanPetPoo:10,HuaSwi:11,LiPetWeb:11}.  Coordinating aggregated interference from multiple LTs at multiple LRs imposes new challenges on secrecy transmission strategy design.
A promising candidate to overcome this challenge is interference alignment (IA) \cite{CadJaf:J08}. A few studies have adopted the IA scheme proposed in \cite{CadJaf:J08} to promote wireless secrecy \cite{KoyGamLaiPoo:11,BasUlu:12,KoyKokElG:12}. However, the scheme in \cite{CadJaf:J08} is based on infinite dimensional symbols that require time or frequency domain symbol extension, making it difficult to implement in practice.

\mysubnote{IA without symbol extension is practical. However, its theoretical foundation is incomplete.}
To avoid the infinite dimension issue, researchers have developed spatial-domain IA techniques, in which no symbol extension is involved and interference is coordinated and canceled via the finite signal dimension provided by multiple antennas \cite{YetGouJaf:10,RuaLauWin:J13}.
In Part I, a theoretical framework has been established to address the two key issues of  spatial-domain IA, i.e., feasibility conditions and transceiver design.
Moreover, to further enhance the network's capability of secrecy protection, legitimate jammers (LJs) are incorporated to better impede ERs without interfering with the LRs.
In this paper, this technique is referred to as GIA.
To apply the GIA technique to practical wireless-tap
networks, the following issues need to be addressed:
\begin{itemize}
\item{\bf Design effective scalable GIA algorithm:} In large-scale networks, the limited policy space in transceiver design is insufficient to cancel interference on all cross links.
Existing works applying IA to large-scale networks \cite{TreCui:10,TreAlfGui:11} address this issue by first dividing a large network into small clusters and then performing IA separately on each cluster.
However, under this approach, the inter-cluster interference is not addressed, and some of it may be the strongest interference perceived by the LRs on a cluster edge.
On the other hand, if every LR wishes to cancel the strongest interference it perceives, the feasibility conditions of the entire network are coupled together, which normally requires centralized approaches that are not applicable to large networks. Hence, designing effective scalable GIA algorithms is difficult.

\item{\bf Characterize the performance of GIA in stochastic networks:}
To obtain insights into the performance of GIA in generic wireless-tap networks, it is desirable to characterize how GIA performs in large-scale stochastic wireless-tap networks.
A few works have analyzed the performance of stochastic networks with interference control \cite{GiaGanHae:11,HuaAndHea:12}.
In these works, the interference control policies at different nodes are independent.
However, with GIA, the interference control policies at different LTs and LJs become correlated, making it difficult to quantify aggregate interference at LRs and ERs.
Therefore, characterizing the performance of GIA in stochastic networks is  challenging.
\end{itemize}
\subsection{Contribution of This Work}
In this work, we will address the challenges listed above.
We consider MIMO wireless-tap networks with LJs.
To enable the design of effective and scalable GIA algorithms,
we first decompose the GIA feasibility conditions to per-node basis.
Based on that, we propose an algorithm that generates a feasible alignment set by only requiring each legitimate node to communicate with a few nodes, the number of which does not scale with the size of the network.
This algorithm, together with the distributive GIA transceiver design algorithm proposed in Part I, construct a GIA algorithm that is applicable to large-scale wireless-tap networks.
We then characterize the performance of the proposed algorithm in stochastic wireless-tap networks.
We jointly adopt Cauchy--Schwarz inequality, Chebyshev inequality, and Chernoff inequality to bound the effect of aggregate interference from multiple correlated sources, and obtain the performance of GIA.
This result demonstrates the contribution of GIA to network secrecy enhancement. It also illustrates how major network parameters, such as node density and antenna configuration, affect the performance of wireless-tap networks.
We also perform various simulations to obtain insights into when and the how GIA technique benefits network secrecy.

\subsection{Organization}
Section~\ref{sec:model} formulates the alignment set design problem.
Section~\ref{sec:alg} proposes a distributive alignment set  design algorithm.
Section~\ref{sec:perf} analyzes the performance of the proposed algorithm in large-scale stochastic wireless-tap networks. Section V provides numerical results. Section~\ref{sec:conclude} concludes.

\mynote{Notation}
\subsection{Notations}
\subsubsection{General}
$a$, $\mathbf{a}$, $\mathbf{A}$, and $\mathcal{A}$ represent scalar, vector, matrix, and set/space, respectively. $\mathbb{N}$, $\mathbb{Z}$, $\mathbb{R}$ and $\mathbb{C}$ denote the set of natural numbers, integers, real numbers, and complex numbers, respectively.

\subsubsection{Functions}
Function $[\cdot]^+=\max\{\cdot,0\}$, $\Gamma(\cdot)$ and $\Gamma(\cdot,\cdot)$ are the gamma function and incomplete gamma function, respectively. $n|m$ denotes that $n$ divides $m$, and $n\;\mathrm{mod}\;m$ denotes $n$ modulo $m$, $n,m\in\mathbb{Z}$.
$\lfloor \cdot \rfloor$ denotes the floor function.
$\mathbb{I}\{\cdot\}$ is the indicator function.
$\begin{pmatrix}n\vspace{-2mm}\\m\end{pmatrix}$ is the Binomial coefficient with parameters $n,m\in\mathbb{N}$. $|a|$ represents the absolute value of scalar $a$, and $|\mathcal{A}|$ represents the cardinality of set $\mathcal{A}$.
\subsubsection{Linear algebra}
The operators $(\cdot)^\mathrm{T}$,  $\bar{(\cdot)}$, $(\cdot)^\mathrm{H}$, $\mathrm{rank}(\cdot)$, $||\cdot||_{\mathrm{F}}$, $\mathrm{trace}(\cdot)$, denote transpose, complex conjugate, Hermitian transpose, rank, Frobenius norm, trace of a matrix. $\mathrm{span}(\mathbf{A})$ and $\mathrm{span}(\{\mathbf{a}\})$ denote the linear space spanned by the column vectors of $\mathbf{A}$ and the vectors in set $\{\mathbf{a}\}$, respectively.
$\dim(\cdot)$ denotes the dimension of a space.
$\mathrm{diag}(\mathbf{A},\ldots,\mathbf{X})$ represents a block diagonal
matrix with submatrices $\mathbf{A},\ldots,\mathbf{X}$.

\subsubsection{Probability theory}
The operators $\mathbb{E}\{\cdot\}$, $\mathbb{V}\{\cdot\}$, and $\mathbb{S}\{\cdot\}$ denote the expectation, variance, and standard deviation of a random variable, and $\Pr\{\cdot\}$ denotes the probability of an event.
$\mathcal{N}_c (\mu, \sigma^2)$ represents complex Gaussian distribution, with mean $\mu$ and standard deviation $\sigma$.

\section{Problem Formulation}

\label{sec:model}
In this section, we will first describe the system model of wireless-tap networks,
then illustrate the potential benefits of GIA via a case study,
and finally formulate the alignment set design problem of GIA.
\subsection{System Model}
\label{subsec:model}
Consider a network consisting of $K$ LT-LR pairs, $J$ LJs and $K$ ERs (The LTs and LJs are indexed from 1 to $K$ and from $K\!+\!1$ to $K\!+\!J$, respectively.). Suppose LT $j$ (or LJ $j$, if $j>K$), LR $k$, and ER $k$ are equipped with $M_j$, $\mnc{N}{\ml}{k}$, and $\mnc{N}{e}{k}$ antennas, respectively. At each time slot, LT (or LJ) $j$ sends $d_j$ independent symbols.
LT $k$ attempts to send confidential messages to LR $k$, while ER $k$ attempts to intercept these messages. LJ~$j$ transmits dummy data to generate interference.

The received signals $\MNC{y}{\ml}{k},\MNC{y}{e}{k}\in \mathbb{C}^{d_k}$ at LR $k$ and ER $k$ are given by
\begin{eqnarray}
\MNC{y}{\iota}{k}=(\MNC{U}{\iota}{k})^\mathrm{H}\bigg(\mathbf{H}^{{(\!\!\;\iota\!\!\;)}}_{kk}
\mathbf{V}_{k}\mathbf{x}_{k} + \sum_{j=1,\neq k}^{\tilde{K}}
\mathbf{H}^{{(\!\!\;\iota\!\!\;)}}_{kj} \mathbf{V}_{j}\mathbf{x}_{j}
+\mathbf{z}^{{(\!\!\;\iota\!\!\;)}}_k\bigg),
\label{eqn:LRsignal}
\end{eqnarray}
where $\tilde{K}=K+J$, $\MNC{H}{\iota}{kj}\in \mathbb{C}^{N^{(\!\!\;\iota\!\!\;)}_k\times M_j}$, ${\iota}\in\{\ell,{e}\}$ are the channel matrices from LT (or LJ) $j$ to LR $k$ or ER $k$, whose entries are independent random variables drawn from continuous distributions; $\mathbf{x}_j \in \mathbb{C}^{d_j}$  is the encoded
information symbol at LT (or LJ) $j$; $\mathbf{V}_{j}\in \mathbb{C}^{M_j\times d_j}$ is the precoder at LT (or LJ) $j$; $\mathbf{U}^{(\!\!\;\iota\!\!\;)}_{k}\in \mathbb{C}^{N^{(\!\!\;\iota\!\!\;)}_k\times d_k}$, ${\iota}\in\{\ell,{e}\}$ is the decoder at LR
$k$ or ER $k$; and $\MNC{z}{\iota}{k}\in \mathbb{C}^{\mnc{N}{\iota}{k}\times 1}$, ${\iota}\in\{\ell,{e}\}$ is the
white Gaussian noise with zero mean and unit variance. The transmission power of LT (or LJ) $j$ is given by $
P_j = \mathbb{E}\left\{\mathrm{trace}\big(\mathbf{x}^\mathrm{H}_j\mathbf{V}^\mathrm{H}_j\mathbf{V}_j\mathbf{x}_j\big) \right\}$.
Define the configuration of the legitimate network as $\chi\triangleq\{(M_1,M_2,\ldots,M_{\tilde{K}}),(\mnc{N}{\ml}{1},\mnc{N}{\ml}{2},\ldots,
\mnc{N}{\ml}{K}),$ $(d_1,d_2,\ldots,d_{\tilde{K}})\}$.

This work adopts Information-theoretic security \cite{LeuHel:78} as the performance metric. From \cite{TanLiuSpaPoo:08,OggHas:11},
under a given transceiver design, the following secrecy rate $R^\mathrm{S}_k$ is achievable for legitimate link $k$:
\begin{eqnarray}
R^\mathrm{S}_k=\mathbb{E}\big\{[\mnc{r}{\ml}{k}-\mnc{r}{e}{k}]^+\big\},\label{eqn:MIMOC}
\end{eqnarray}
in which $\mnc{r}{\iota}{k}$, $\iota\in\{\ell,{e}\}$, is given by
\begin{eqnarray}
\mnc{r}{\iota}{k}=\log_2\det\bigg\{\!\mathbf{I}\!+
\!(\MNC{U}{\iota}{k})^\mathrm{H}
\MNC{H}{\iota}{kk}\mathbf{V}_{k}
\big(\MNC{H}{\iota}{kk}\mathbf{V}_{k}\big)^\mathrm{H}
\MNC{U}{\iota}{k}\nonumber\hspace{5mm}
\\
\Big[(\MNC{U}{\iota}{k})^\mathrm{H}\Big(\mathbf{I}\!
+\!\sum_{j=1,\neq k}^{\tilde{K}}
\MNC{H}{\iota}{kj} \mathbf{V}_{j}\big(\MNC{H}{\iota}{kj} \mathbf{V}_{j}\big)^\mathrm{H}\Big)\MNC{U}{\iota}{k}\Big]^{\!-\!1}
\!\bigg\}.\label{eqn:rate_LR}
\end{eqnarray}

From \cite{BasUlu:12},
when the transmission power at all LTs and LJs are on the same order, i.e., for some  $\theta_{\mathrm{l}},\theta_{\mathrm{h}}>0$, $\frac{P_j}{P}\in[\theta_{\mathrm{l}},\theta_{\mathrm{h}}]$, $\forall j$, the secure degree of freedom (sDoF) can be defined as
\begin{eqnarray}
D^\mathrm{S}_k =\lim_{P\rightarrow\infty}\frac{R^\mathrm{S}_k}{\log_2(P)}. \label{eqn:sDoF}
\end{eqnarray}

\subsection{Case Study}
\label{subsec:case}
\begin{figure}[t] \centering
\includegraphics[scale=0.7]{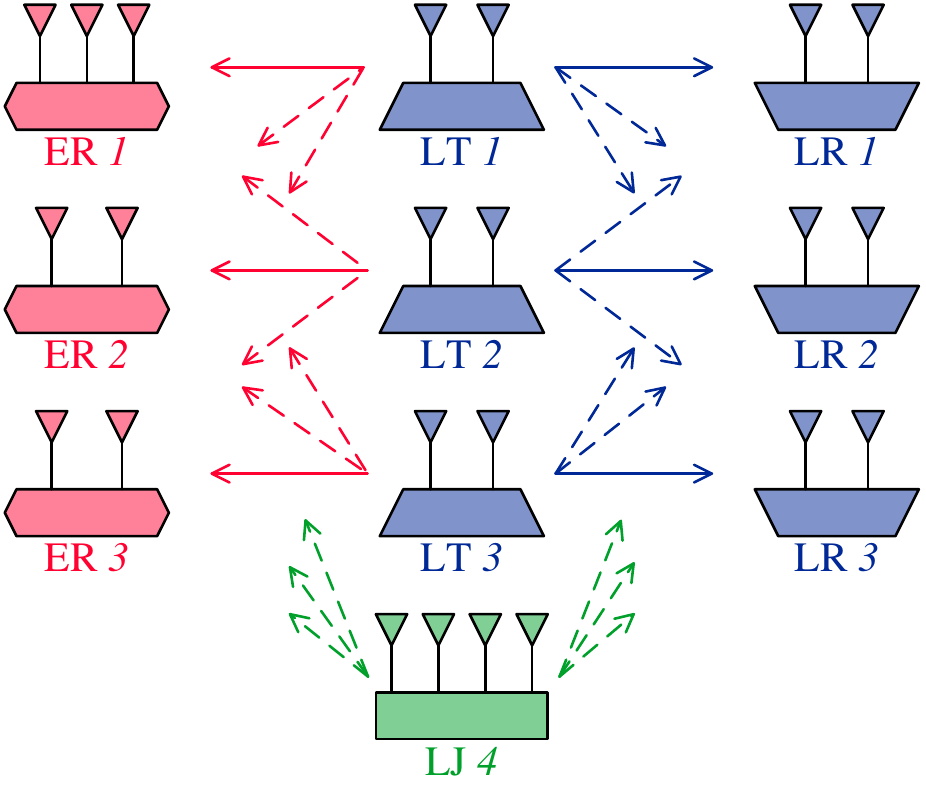}
\caption{Configuration of the example network.}
\label{fig_MotiExp}
\end{figure}

\emph{Example:} As illustrated in Fig.~\ref{fig_MotiExp}, consider a MIMO wireless-tap network, as described in Sec.~\ref{subsec:model}, with $K=3$, $J=1$, antenna configuration $M_k=\mnc{N}{\ml}{k}=2$, $k\in\{1,2,3\}$,  $M_4=4$, $\mc{N}{e}_1=3$, $\mc{N}{e}_2=\mc{N}{e}_3=2$, and data stream configuration $d_k=1$, $k\in\{1,2,3,4\}$. The entries of all the channel matrices are independent random variables drawn from $\mathcal{N}_c(0,1)$. The transmit power at all nodes is 20dB, i.e., $P_k=P=100$, $k\in\{1,2,3,4\}$. ~\hfill~\IEEEQED

Compare four different strategies: the first two are non-cooperative, whereas the other two are cooperative:
\begin{itemize}
\item[\bf A.]\savespacess{\em Zero forcing with 2 active LTs:} LT 1, 2 use random precoders $\mathbf{v}_1$, $\mathbf{v}_2$,\footnote{Small letters are used for all transceivers as they are vectors in this example.} and LR 1, 2 use zero forcing to cancel  interference, i.e., $\mc{\mathbf{u}}{{\ml}}_1=
\mathbf{Z}\overline{\mc{\mathbf{H}}{{\ml}}_{12}{\mathbf{v}}_2}$ and
$\mc{\mathbf{u}}{{\ml}}_2
=\mathbf{Z}\overline{\mc{\mathbf{H}}{{\ml}}_{21}\mathbf{v}_1}$, where
$\mathbf{Z}=\bigg[\!\!\begin{array}{c@{\;\;}c}0&-\!1\\1&0\end{array}\!\!\bigg]$. To avoid interfering with LR 1 and 2, LT 3 and LJ 4 remain silent, i.e., $\mathbf{v}_3=\mathbf{0}$, $\mathbf{v}_4=\mathbf{0}$.
\item[\bf B.]\savespacess{\em Zero forcing with 3 active LTs:} LT 1, 2 and 3 use random precoders $\mathbf{v}_1$, $\mathbf{v}_2$ and $\mathbf{v}_3$, and LR 1, 2 and 3 use zero forcing to cancel interference from LT 2, 3 and 1, respectively, i.e., $\mc{\mathbf{u}}{{\ml}}_1=\mathbf{Z}\overline{\mc{\mathbf{H}}{{\ml}}_{12}\mathbf{v}_2}$, $\mc{\mathbf{u}}{{\ml}}_2=\mathbf{Z}\overline{\mc{\mathbf{H}}{{\ml}}_{23}\mathbf{v}_3}$ and $\mc{\mathbf{u}}{{\ml}}_3=\mathbf{Z}\overline{\mc{\mathbf{H}}{{\ml}}_{31}\mathbf{v}_1}$. LJ 4 remains silent.
\item[\bf C.]\savespacess{\em IA:} LT 1--3 adopt IA to design precoders $\{\mathbf{v}_k\}$, $k\in\{1,2,3\}$. Interference at every LR is aligned into a 1-dimensional subspace. Specifically, $\mathbf{v}_1$ is designed to be a eigenvector of
$(\mc{\mathbf{H}}{{\ml}}_{31})^{-\!1}\mc{\mathbf{H}}{{\ml}}_{32}(\mc{\mathbf{H}}{{\ml}}_{12})^{-\!1}
    \mc{\mathbf{H}}{{\ml}}_{13}(\mc{\mathbf{H}}{{\ml}}_{23})^{-\!1}\mc{\mathbf{H}}{{\ml}}_{21}$, $\mathbf{v}_2=(\mc{\mathbf{H}}{{\ml}}_{32})^{-\!1}\mc{\mathbf{H}}{{\ml}}_{31}\mathbf{v}_1$, and $\mathbf{v}_3=(\mc{\mathbf{H}}{{\ml}}_{23})^{-\!1}\mc{\mathbf{H}}{{\ml}}_{21}\mathbf{v}_1$. $\mc{\mathbf{u}}{{\ml}}_1$, $\mc{\mathbf{u}}{{\ml}}_2$, and $\mc{\mathbf{u}}{{\ml}}_3$ are designed as in Strategy B. LJ 4 still remains silent.
\item[\bf D.]{\em GIA:} LT 1--3 and LJ 4 adopt a coordinated approach to design their precoders so that interference at every LR is aligned to a 1-dimensional subspace. Specifically, the LTs and LRs design their transceivers as in Strategy C.
LJ 4 designs $\mathbf{v}_4$ so that $\mathbf{v}_4\perp\big((\mc{\mathbf{u}}{{\ml}}_k)^\mathrm{H}\mc{\mathbf{H}}{{\ml}}_{k4}\big)^{\mathrm T}$,  $k\in\{1,2,3\}$.
This design is feasible as $\big((\mc{\mathbf{u}}{{\ml}}_k)^\mathrm{H}\mc{\mathbf{H}}{{\ml}}_{k4}\big)^{\mathrm
T}$,  $k\in\{1,2,3\}$ are three vectors in $\mathbb{C}^4$.
\end{itemize}

\begin{figure}[t]
\hspace{-6mm}\includegraphics[scale=0.57]{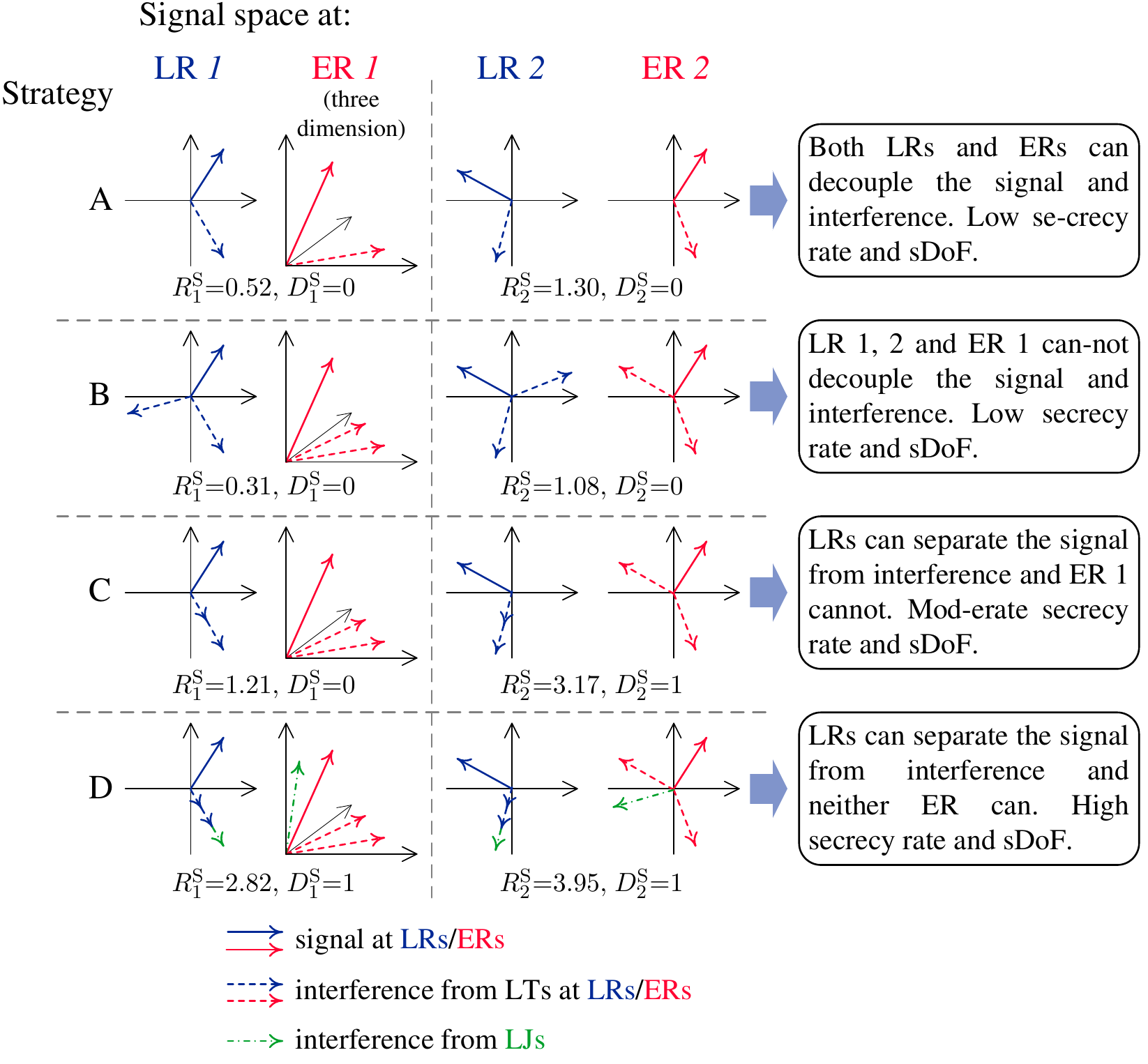}
\caption{Signal space at LR 1, 2 and ER 1, 2 under different strategies.}
\label{fig_ExpSig}
\end{figure}

The signal space at LR 1, 2 and ER 1, 2 as well as the secrecy rate and sDoF of LR 1, 2 under the above four strategies are illustrated and compared in Fig.~\ref{fig_ExpSig}.\footnote{Because the cases of LR 3/ER 3 are identical to those of LR 2/ER 2 under Strategy B-D, they are omitted in Fig.~\ref{fig_ExpSig} for conciseness.} From this figure, Strategy C, D perform better than Strategy A, B. This is because the channel states between LTs and LRs $\{\mc{\mathbf{H}}{\ml}_{kj}\}$
are independent of those between LTs and ERs $\{\mc{\mathbf{H}}{e}_{kj}\}$. Therefore, interference that is aligned at the LRs is not aligned at the ERs almost surely. This fact allow the legitimate network to impede ERs without affecting LRs. Strategy D performs best as its jointly exploits the capability of all legitimate partners, i.e., LTs, LRs, and LJs to generate desirable interference.

\begin{Remark}[Practical Issues] It is worth noting that the GIA technique proposed in the example:
\begin{itemize}
\item{\bf requires no channel state for the eavesdropping link.} The transceivers of the legitimate partners are functions of the channel state  of legitimate links, i.e., $\{\mc{\mathbf{H}}{\ml}_{kj}\}$. In other words, no channel state information (CSI) of the eavesdropping link, i.e., $\{\mc{\mathbf{H}}{e}_{kj}\}$ is required.
\item{\bf is effective even if ERs have all the CSI.} The performance of Strategy D comes from the unequal dimension of the interference at the ERs and LRs. Since this property is due to the fact that the two sets of channel state $\{\mc{\mathbf{H}}{\ml}_{kj}\}$ and $\{\mc{\mathbf{H}}{e}_{kj}\}$ are independent, it is invariant with respect to the amount of CSI at the ERs.
\end{itemize}
These properties greatly improves the practicality of the proposed GIA technique.
For instance, the possible leakage of CSI from the legitimate network to ERs does not affect the performance of the proposed algorithms.~\hfill~\IEEEQED
\end{Remark}

\subsection{Alignment Set Design}

In Sec.~\ref{subsec:case}, the potential of GIA technique  in secrecy enhancement is demostrated.
To cope with the general cases, the following problem is addressed in Part I :
\begin{Prob}[GIA Transceiver Design]\label{pro:GIA}
Design transceivers $\{\MNC{U}{\ml}{k},$ $ \mathbf{V}_j\}$, $k\in\{1,\ldots,K\}$, $j\in\{1,\ldots,\tilde{K}\}$ that satisfy the following constraints:
\begin{eqnarray}
\mathrm{rank}\left((\MNC{U}{\ml}{k})^\mathrm{H}\MNC{H}{\ml}{kk}\mathbf{V}_k\right) \!\!&=\!\!& d_k\label{eqn:drank}, \quad\forall k\in\{1,\ldots,K\},\\
\mathrm{rank}\left(\mathbf{V}_j\right) \!\!&=\!\!& d_j\label{eqn:rankJ}, \quad\forall j\in\{K\hspace{-0.7mm}+\!1,\ldots,\tilde{K}\},\\
\mbox{ and}\quad(\MNC{U}{\ml}{k})^\mathrm{H}\MNC{H}{\ml}{kj}\mathbf{V}_j \!\!&=\!\!& \mathbf{0},\quad\hspace{1.2mm}\forall (k,j)\in \mathcal{A},\label{eqn:czero}
\end{eqnarray}
where $\mathcal{A}\subseteq\mathcal{A}_{\mathrm{all}}=\{(k,j):\; k\in\{1,\ldots,K\},j\in\{1,\ldots,\tilde{K}\},k\neq j\}$ is the \emph{alignment set}.
It characterizes the set of interference to be canceled by GIA.~\hfill~\IEEEQED
\end{Prob}

In Part I, the feasibility conditions of Problem~\ref{pro:GIA} is analyzed for given network configuration $\chi$ and alignment set $\mathcal{A}$. In practice, the network configuration is usually fixed at prior. Hence, to design feasible GIA strategies, the following problem needs to be addressed:
\begin{Prob}[Alignment Set Design]\label{pro:Set}Design $\mathcal{A}$ so that GIA\ is feasible, i.e., Problem~\ref{pro:GIA} has solutions.~\hfill~\IEEEQED
\end{Prob}

To develop GIA techniques that are applicable to large-scale networks, it is important to design algorithms that can solve Problem~\ref{pro:Set} distributively. However, this task is difficult due to the the following technical challenge.

\vspace{3mm}\noindent
\framebox{
\parbox{8.3cm}{
\textbf{Challenge of Coupled Feasibility Conditions}

As Corollary~4.3 of Part I shows, for GIA to be feasible, it is necessary that the number of variables in transceiver design is no less than the number of constraints for {\em all} subsets of GIA constraints in \eqref{eqn:czero}. This fact illustrates that GIA feasibility conditions are inherently coupled with each other.
Since there are exponentially many subsets of GIA constraints, the design of
a feasible alignment set is complicated.
}}
\vspace{3mm}

\section{Algorithm Design}
\label{sec:alg}
In this section, a GIA algorithm will be proposed to solve Problem~\ref{pro:Set} distributively. To achieve this, we first decompose the GIA feasibility conditions to per-node basis via Theorem~\ref{thm:cover}.

\begin{Def}[Proper Alignment Subsets]\label{def:proper} Alignment subsets $\mc{\mathcal{A}}{r}(k)=\{(k,j)\}\subseteq \{k\}\times\big(\{1,\ldots,\tilde{K}\}\backslash\{k\}\big)$, and $\mc{\mathcal{A}}{t}(j)=\{(k,j)\}\subseteq \big(\{1,\ldots,K\}\backslash\{j\}\big) \times \{j\}$ are \emph{proper} iff.
\begin{eqnarray}
\sum_{j:(k,j)\in\atop \mc{\mathcal{A}}{r}(k)} d_j \le \mnc{N}{\ml}{k} -d_k,\quad  \sum_{k:(k,j)\in\atop \mc{\mathcal{A}}{t}(j)} d_k \le M_j -d_j. \label{eqn:proper}
\end{eqnarray}~\hfill~\IEEEQED
\end{Def}

\begin{Thm}[Proper Alignment Subsets Lead to GIA Feasibility]\label{thm:cover} Problem~\ref{pro:GIA} is feasible almost surely if the alignment set $\mathcal{A}$ can be covered by proper alignment subsets, i.e.,
\begin{eqnarray}
\mathcal{A} = \Big( \bigcup_{k\in\{1,\ldots,K\}} \mc{\mathcal{A}}{r}(k) \Big)\bigcup \Big(\bigcup_{j\in\{1,\ldots,\tilde{K}\}} \mc{\mathcal{A}}{t}(j)\Big)\label{eqn:cover}
\end{eqnarray}
for some proper alignment subsets $\mc{\mathcal{A}}{r}(k)$ and $\mc{\mathcal{A}}{t}(j)$.
\end{Thm}
\begin{IEEEproof} Please refer to Appendix~\ref{pf_thm:cover} for the proof.
\end{IEEEproof}

\vspace{3mm}\noindent
\framebox{
\parbox{8.3cm}{
\textbf{Solution to Coupled Feasibility Conditions}

Since equation \eqref{eqn:proper} is a set of
per-node constraints,
Theorem~\ref{thm:cover} provides a mechanism to decompose the GIA feasibility constraints to per-node basis.
This result enables legitimate nodes to distributively design the alignment set, while maintaining the GIA feasibility.
}}
\vspace{3mm}

Based on Theorem~\ref{thm:cover}, the following algorithm is adopted to
generate alignment set $\mathcal{A}$.
\begin{Alg}[Generate Feasible Alignment Set]
\label{alg:alignmentset}
\begin{itemize}
\item {\bf Alignment set selection at the transmitter side:}
LT (or LJ) $j$ selects a few LRs  such that $\mc{\mathcal{A}}{t}(j)$ satisfies \eqref{eqn:proper}.\footnote{Here node selection criteria is not specified as it does not affect the feasibility of the alignment
set. The selection criteria will be specified in the next section to enable performance analysis. } Notify the selected LRs.
\item {\bf Alignment set selection at the receiver side:} LR $k$ selects among the transmitters which do not select LR $k$ in the previous step, and make $\mc{\mathcal{A}}{r}(k)$ satisfy
\eqref{eqn:proper}.
\item {\bf Generate Alignment Set:} Set $\mathcal{A}$ according to \eqref{eqn:cover}.
~\hfill~\IEEEQED
\end{itemize}
\end{Alg}

\begin{Cor}[Feasibility of Algorithm~\ref{alg:alignmentset}] In a MIMO wireless-tap network, when the alignment set $\mathcal{A}$
is generated by Algorithm~\ref{alg:alignmentset}, Problem~\ref{pro:GIA} is feasible almost
surely.
\end{Cor}
\begin{IEEEproof} This corollary is a direct consequence of Theorem~\ref{thm:cover}.
\end{IEEEproof}

\begin{Remark}[Effective Scalable GIA Algorithm]
The freedom in designing the alignment subsets,  $\mc{\mathcal{A}}{t}(j)$, $\mc{\mathcal{A}}{r}(k)$ in Algorithm~\ref{alg:alignmentset} enables the legitimate nodes to distributively select the strongest interfering links and hence effectively manage interference.
Since the GIA transceiver design algorithm proposed in Part I can also be executed distributively (please refer to Remark~4.8 of Part I for details), by first performing Algorithm~\ref{alg:alignmentset} to design a feasible alignment set $\mathcal{A}$ and then using the algorithm proposed in Part I to design the transceivers $\{\MNC{U}{\ml}{k},\mathbf{V}_j\}$, a distributive GIA algorithm is obtained.
In this algorithm, the number of nodes that each node need to exchange messages with are determined by the alignment subsets and hence does not scale with the size of the network.~\hfill~\IEEEQED
\end{Remark}

\section{Performance Analysis}
\label{sec:perf}

In this section, we will check how the insight in Sec.~\ref{subsec:case} applies to generic wireless-tap networks.

In today's wireless networks, the randomness in aggregated interference is largely attributed to the locations of active interferers \cite{MorLoy:09}.
Consequently, to characterize the performance of wireless networks, researchers have modeled the locations of nodes using random point processes.
To maintain analytical tractability, homogeneous point processes are usually adopted.
In particular, the homogeneous Poisson point process (PPP)  is widely adopted  \cite{GovBliSta:07,HaeAndBacDouFra:09} as it possesses the highest entropy and accounts for  complete randomness in node locations.
The PPP has been used to analyze the connectivity of wireless networks with secrecy.
In this work, PPP is also adopted for the framework of GIA performance evaluation.

\begin{Def}[Stochastic Wireless-tap Network]\label{def:typology}
\begin{itemize}
\item{\em Channel Model:} The nodes are distributed in a two-dimensional infinite plane $\mathbb{R}^2$. The channel state between two nodes positioned at $\mathbf{a}, \mathbf{b} \in\mathbb{R}^2$ is given by $\mathbf{H}_{\mathbf{a},\mathbf{b}}=L(\mathbf{a},\mathbf{b})\tilde{\mathbf{H}}_{\mathbf{a},\mathbf{b}}$, where the elements in $\tilde{\mathbf{H}}$ are independent random variables following complex Gaussian distribution with zero mean and unit variance and the pathloss
    \begin{eqnarray}
    L(\mathbf{a},\mathbf{b}) = ||\mathbf{a}-\mathbf{b}||^{-\frac{\alpha}{2}}
    \mathbb{I}\{||\mathbf{a}-\mathbf{b}||^{-\frac{\alpha}{2}}\ge \theta\}\label{eqn:pathloss},
    \end{eqnarray}
    where $\alpha\in[2,4]$ is the pathloss exponent and $\theta$ is the cutoff threshold.\footnote{Suppose the maximum transmit power of nodes in the network is $P_{\mathrm{max}}$. Then when $\theta^2 P_{\mathrm{max}}\ll 1$, the interference that has been ignored by the pathloss cutoff threshold is insignificant compared to white noise. In this case, the pathloss model in \eqref{eqn:pathloss} is a reasonable approximation of the classical one.
The two models will be compared via simulation in Fig.~\ref{fig_throughput}.}
\item{\em Legitimate user network:} The position of the LTs is modeled by a homogeneous PPP with density $\mc{\lambda}{\ml}$. For an LT located at $\mathbf{b}$,
the position of the associated LR is given by $\mathbf{a}=\mathbf{b}+\mathbf{p}_{\mathbf{b}}$, where $\mathbf{p}_{\mathbf{b}}$ is drawn from certain probability distribution in $\mathbb{R}^2$, with $||\mathbf{p}_{\mathbf{b}}||\le \theta^{-\frac{2}{\alpha}}$.\footnote{Otherwise, from \eqref{eqn:pathloss},
the channel between the LR and the associated  LT  is $\mathbf{0}$, which
leads to trivial result. For the same reason, the distance between LT and the corresponding ER is limited.} Each LR and LT is equipped with $\mc{M}{\ml}$ and $\mc{N}{\ml}$ number of antennas, respectively. Each LT delivers $\mc{d}{\ml}$ ($\le\min\{\mc{M}{\ml}$, $\mc{N}{\ml}\}$) independent data streams. Denote $\mathbf{p}_\mathbf{a}=-\mathbf{p}_{\mathbf{b}}$.
\item{\em Legitimate jammer network:} The position of the LJs is modeled by a PPP with density $\mc{\lambda}{j}$. Each LJ has $\mc{M}{j}$  number of antennas and delivers $\mc{d}{j}$ ($\le\mc{M}{j}$) independent dummy data streams.
\item{\em Eavesdropper network:} The position of the ER attempting to intercept the information from the LT at position $\mathbf{b}$ is given by $\mathbf{e}=
\mathbf{b}+\tilde{\mathbf{p}}_{\mathbf{b}}$, where $\tilde{\mathbf{p}}_{\mathbf{b}}$ is drawn from certain probability distribution in $\mathbb{R}^2$, with $||\tilde{\mathbf{p}}_{\mathbf{b}}||\le \theta^{-\frac{2}{\alpha}}$.
Each ER is equipped with $\mc{N}{e}$ number of antennas and adopts minimum mean square error decoder. Denote
$\mathbf{p}_\mathbf{e}=-\tilde{\mathbf{p}}_{\mathbf{b}}$.~\hfill~\IEEEQED
\end{itemize}
\end{Def}

For notation convenience, in the following, the position of a node will be used to replace its index. For example, an LT positioned at $\mathbf{b}\in\mathbb{R}^2$ is denoted by LT $\mathbf{b}$.
Denote the set of the positions of LTs, LRs, LJs, and ERs by $\mc{\mathcal{T}}{\ml}$, $\mc{\mathcal{R}}{\ml}$, $\mc{\mathcal{T}}{j}$, and $\mc{\mathcal{R}}{e}$, respectively.

To cancel the strongest interference that each LR perceives, we set the selection criteria in Algorithm~\ref{alg:alignmentset} so that the nodes select the nearest neighboring nodes first, i.e.,
\begin{itemize}
\item{\bf Transmitter side:} LT $\mathbf{b}$ sets $\mc{\mathcal{A}}{t}(\mathbf{b})
=\{(\mathbf{a},\mathbf{b})\}$, where $\mathbf{a}\in\mc{\mathcal{R}}{\ml}\backslash
\{\mathbf{b}+\mathbf{p}_{\mathbf{b}}\}$, so that
\begin{align}
&\hspace{-7mm}L(\mathbf{a},\mathbf{b})\ge
L(\tilde{\mathbf{a}},\mathbf{b}),\quad \quad\forall \tilde{\mathbf{a}}\in \mc{\mathcal{R}}{\ml}\backslash\nonumber
\\ &
\big(\{\mathbf{a}:(\mathbf{a},\mathbf{b})\in\mc{\mathcal{A}}{t}(\mathbf{b})\}
\cup\{\mathbf{b}+\mathbf{p}_{\mathbf{b}}\}\big),\label{eqn:nearR}
\\
&\hspace{-7mm}\big|\mc{\mathcal{A}}{t}(\mathbf{b})\big|=
\left\lfloor\frac{\mc{M}{x}-\mc{d}{x}}{\mc{d}{\ml}}
\right\rfloor\triangleq\mc{m}{x}\label{eqn:ml},
\end{align}
where $x = \ml, j$ for the LTs and LJs, respectively.
\item{\bf Receiver side:} LR $\mathbf{a}$ sets $\mc{\mathcal{A}}{r}(\mathbf{a})=
\{(\mathbf{a},\mc{\mathbf{b}}{\ml})\}\cup
\{(\mathbf{a},\mc{\mathbf{b}}{j})\}$, where $\mc{\mathbf{b}}{\ml}\in \mc{\mathcal{T}}{\ml}
\backslash\{\mathbf{a}+\mathbf{p}_{\mathbf{a}}\}$,
$\mc{\mathbf{b}}{j}\in \mc{\mathcal{T}}{j}$,   so that
\begin{align}
&\hspace{-3.5mm}(\mathbf{a},\mc{\mathbf{b}}{x})\not\in \mc{\mathcal{A}}{t}(\mc{\mathbf{b}}{x}), \quad x\in\{\ml,j\},\label{eqn:setnooverlap}
\\
&\hspace{-3.5mm}L(\mathbf{a}_,\mc{\mathbf{b}}{x})\ge
L(\mathbf{a},\mathbf{b}),\quad \forall \mathbf{b}\in \big(\mc{\mathcal{T}}{\ml}\cup \mc{\mathcal{T}}{j}\big)\backslash\nonumber
\\&\hspace{-1mm}
\big(\{\mathbf{b}:(\mathbf{a},\mathbf{b})\in\mc{\mathcal{A}}{r}(\mathbf{a})\cup\mc{\mathcal{A}}{t}
(\mathbf{b})\}\cup\{\mathbf{a}+\mathbf{p}_{\mathbf{a}}\}\big),\label{eqn:nearT}
\\
&\hspace{-3.5mm}\mc{N}{\ml} - \mc{d}{\ml} -\max\{\mc{d}{\ml}, \mc{d}{j} \}+1
\le\nonumber
\\
&\hspace{-1mm}\mc{d}{\ml}
\big|\{(\mathbf{a},\mc{\mathbf{b}}{\ml})\}\big|
+\mc{d}{j}\big|\{(\mathbf{a},\mc{\mathbf{b}}{j})\}\big|\le \mc{N}{\ml} - \mc{d}{\ml}.\label{eqn:Set_r}
\end{align}
\end{itemize}

Define the  \emph{connection density} of the legitimate network and the jammer network $\mc{\rho}{\ml}$, $\mc{\rho}{j}$, as the expected number of LTs or LJs that may interfere with a receiver, i.e.,
\begin{eqnarray}
\mc{\rho}{\ml}&\skiph=\skiph&\mathbb{E}\left\{\sum_{\mathbf{b}\in
\mc{\mathcal{T}}{\ml}}
\mathbb{I}\{L(\mathbf{a},\mathbf{b})>0\}\right\}=\pi\theta^{-\frac{4}{\alpha}}\mc{\lambda}{\ml},
\label{eqn:rho_l} \\\mc{\rho}{j}&\skiph=\skiph&\mathbb{E}\left\{\sum_{\mathbf{b}\in\mc{\mathcal{T}}{j}}
\mathbb{I}\{L(\mathbf{a},\mathbf{b})>0\}\right\}=\pi\theta^{-\frac{4}{\alpha}}\mc{\lambda}{j},
\label{eqn:rho_j}
\end{eqnarray}
where $\mathbf{a}\in \mathbb{R}^2$.

This section focuses on analyzing the sDoF achieved by an LR. Firstly, a lemma which relates the sDoF to the dimension of interference at the LRs and ERs is proved.
\begin{Lem}[SDoF and Dimension of Interference]\label{lem:SDI} For LR $\mathbf{a}$ with corresponding ER $\mathbf{e}$, the sDoF defined in \eqref{eqn:sDoF} is given by
\begin{align}
D^\mathrm{S} = \Big[\mathrm{dim}\big(\mc{\mathcal{S}}{\ml}\big)-
\mathrm{dim}\big(\mc{\mathcal{S}}{e}\big)\Big]^+\label{eqn:sDoF2},
\end{align}
where $\mc{\mathcal{S}}{\iota}$, $\iota\in\{\ml,e\}$ is a subspace of the receiving signal space of LR or ER, i.e.,
$\mathrm{span}\big(\MNC{U}{\ml}{\mathbf{a}}\big)$ or $\mathrm{span}\big(\MNC{U}{e}{\mathbf{e}}\big)$ that has no interference. Furthermore, define $\mc{S}{\iota}\triangleq\mathrm{dim}\big(\mc{\mathcal{S}}{\iota}\big)$, $\iota\in\{\ml,e\}$, then
\begin{align}
\mc{S}{\ml} = &\Big[\mc{d}{\ml}- \mc{d}{\ml}\hspace{-2mm}
\sum_{{\mathbf{b}\in\mc{\mathcal{T}}{\ml}}
}\hspace{-1mm} \mathbb{I}\{L(\mathbf{a},\mathbf{b})>0\;\&\;(\mathbf{a},\mathbf{b})\not\in \mathcal{A}\}-
\nonumber
\\& \mc{d}{j}\hspace{-2mm}\sum_{{\mathbf{b}\in\mc{\mathcal{T}}{j}}}\hspace{-1mm} \mathbb{I}\{L(\mathbf{a},\mathbf{b})>0\;\&\;(\mathbf{a},\mathbf{b})\not\in
\mathcal{A}\}\Big]^+,\label{eqn:Sl}\\
\mc{S}{e}=
&\min\bigg\{\mc{d}{\ml},\Big[\mc{N}{e}-
\mc{d}{\ml}\hspace{-2mm}\sum_{{\mathbf{b}\in\mc{\mathcal{T}}{\ml}}}\hspace{-1mm} \mathbb{I}\{L(\mathbf{e},\mathbf{b})>0\}-
\nonumber
\\& \mc{d}{j}\hspace{-2mm}\sum_{{\mathbf{b}\in\mc{\mathcal{T}}{j}}}\hspace{-1mm} \mathbb{I}\{L(\mathbf{e},\mathbf{b})>0\}\Big]^+\bigg\}.
\label{eqn:Se}
\end{align}
\end{Lem}
\begin{IEEEproof}
Please refer to Appendix~\ref{pf_lem:SDI} for the proof.
\end{IEEEproof}

From Lemma~\ref{lem:SDI}, to analyze the network's sDoF, the characterization of $\mc{S}{\ml}$ is necessary.
However, this is challenging for the following reason.

\begin{figure}[t] \centering
\includegraphics[scale=0.85]{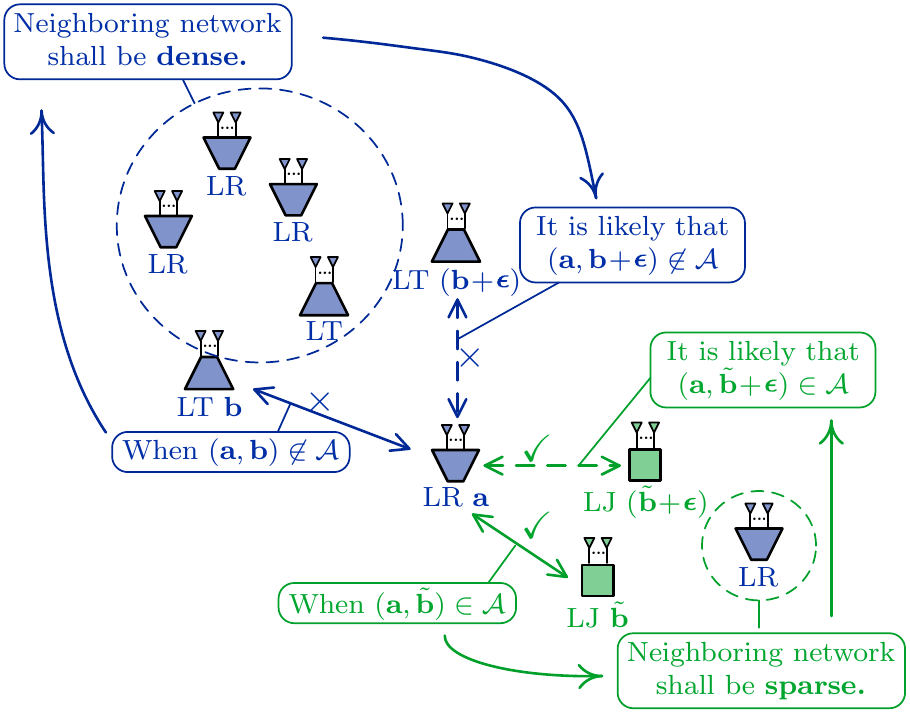}
\caption {Illustration of the correlation between alignment subsets of neighboring
nodes. Consider two LTs (or LJs) positioned at $\mathbf{b}$ and $\mathbf{b}+\boldsymbol{\epsilon}$,
respectively, where
$\boldsymbol{\epsilon}\in\mathbb{R}^2$ has a small norm. Events $(\mathbf{a},\mathbf{b})\in\mathcal{A}$
and $(\mathbf{a},\mathbf{b}+\boldsymbol{\epsilon})\in\mathcal{A}$ are correlated
as the two transmitters perceive similar neighboring networks. } \label{fig_correlation}\vspace{-3mm}
\end{figure}

\vspace{3mm}\noindent
\framebox{
\parbox{8.3cm}{
\textbf{Challenge of Correlated Alignment Set Selection}

As illustrated in Fig.~\ref{fig_correlation}, the events that the links between one LR and several neighboring transmitters being in the alignment set $\mathcal{A}$, e.g., events $(\mathbf{a},\mathbf{b})\in\mathcal{A}$ and $(\mathbf{a},\mathbf{b}+\boldsymbol{\epsilon})\in\mathcal{A}$ in the figure, are correlated.
This fact makes the random variables $\mathbb{I}\{L(\mathbf{a},\mathbf{b})>0\;\&\;(\mathbf{a},\mathbf{b})\not\in
\mathcal{A}\}$ in \eqref{eqn:Sl} correlated and hence makes it difficult to characterize $\mc{S}{\ml}$.
}}
\vspace{3mm}

To overcome this challenge, Cauchy--Schwarz inequality and Chernoff inequality are adopted to bound the effect of correlated alignment set selection. This result is summarized by the following lemma.

\begin{Lem}[Characterization of $\mc{S}{\ml}$]
For LR $\mathbf{a}$, define
\begin{align}
\mc{I}{\ml}\triangleq&\,\mc{d}{\ml}\hspace{-5mm}
\sum_{{\mathbf{b}\in\mc{\mathcal{T}}{\ml}\atop (\mathbf{a},\mathbf{b})\not\in
\mc{\mathcal{A}}{\ml}(\mathbf{b})}
}\hspace{-4mm} \mathbb{I}\{L(\mathbf{a},\mathbf{b})>0\}+\mc{d}{j}
\hspace{-5mm}\sum_{{\mathbf{b}\in\mc{\mathcal{T}}{j}}\atop(\mathbf{a},\mathbf{b})\not\in
\mc{\mathcal{A}}{j}(\mathbf{b})}\hspace{-4mm}
\mathbb{I}\{L(\mathbf{a},\mathbf{b})>0\},\label{eqn:Il}\\
\varepsilon\triangleq&\,\mc{N}{\ml}-\mc{d}{\ml}-\mc{d}{\ml}\big|\{\mathbf{b}\in\mc{\mathcal{T}}{\ml}:(\mathbf{a},\mathbf{b})
\in\mc{\mathcal{A}}{r}(\mathbf{a})\}\big|\nonumber
\\&-\mc{d}{j}\big|\{\mathbf{b}\in\mc{\mathcal{T}}{j}:(\mathbf{a},\mathbf{b})
\in\mc{\mathcal{A}}{r}(\mathbf{a})\}\big|\label{eqn:epsilon}
\end{align}
Then
\begin{eqnarray}
\mc{S}{\ml} =
\min\Big\{\mc{d}{\ml},\big[\mc{N}{\ml}-\mc{I}{\ml}-\varepsilon\big]^+
\Big\}.\label{eqn:Sl2}
\end{eqnarray}
Moreover, $\varepsilon$ is bounded within $\Big[0,\max\{\mc{d}{j},\mc{d}{\ml}\}-1\Big]$,
\begin{align}
\hspace{-3mm}\mathbb{E}\{\mc{I}{\ml}\}\in&
\bigg[\sum_{x\in\{\ml,j\}}
\Big[\mc{\rho}{x}\mc{d}{x}-\frac{\mc{\lambda}{x}}{\mc{\lambda}{l}}\mc{m}{x}\mc{d}{x}\Big]^+,\sum_{x\in\{\ml,j\}}\nonumber
\\&\Big[\mc{\rho}{x}\mc{d}{x}-\frac{\mc{\lambda}{x}}{\mc{\lambda}{l}}\mc{m}{x}\mc{d}{x}\Big]^+
+\frac{\mc{d}{x}\mc{\lambda}{x}\sqrt{\mc{\rho}{\ml}}}{\mc{\lambda}{\ml}
\sqrt{2\pi}}\bigg], \label{eqn:eIl}\\
\hspace{-3mm}\mathbb{S}\{\mc{I}{\ml}\}\le&\sum_{x\in\{\ml,j\}}\frac{4\mc{\lambda}{x}\sqrt{\pi\min\{\mc{m}{x},\mc{\rho}{\ml}
\}}}{\mc{\lambda}{l}}\nonumber
\\&(1+\frac{1}{6\min\{\mc{m}{x},\mc{\rho}{\ml} \}}).
\label{eqn:vIl}
\end{align}
\label{lem:dim_l}
\end{Lem}
\begin{IEEEproof} Please refer to Appendix~\ref{pf_lem:dim_l} for the proof.
\end{IEEEproof}

\vspace{3mm}\noindent
\framebox{
\parbox{8.3cm}{
\textbf{Solution to Correlated Alignment Set Selection}

From \eqref{eqn:Sl2}, the major randomness of $\mc{\mathcal{S}}{\ml}$ comes from that of $\mc{I}{\ml}$.
Equations \eqref{eqn:eIl} and \eqref{eqn:vIl} show that the expectation of
$\mc{I}{\ml}$ scales at $\mathcal{O}(\mc{\rho}{\ml})$, while its uncertainty in expectation and standard
deviation both scale at $\mathcal{O}(\sqrt{\mc{\rho}{\ml}})$.
Therefore, when $\mc{\rho}{\ml}$ is large, the randomness in $\mc{I}{\ml}$
is ignorable compared to its expectation, i.e., $\lim_{\mc{\rho}{\ml}\rightarrow
+\infty}\frac{\mc{I}{\ml}}{\mathbb{E}\{\mc{I}{\ml}\}}= 1$.
As will be further discussed in Remark~\ref{remark:sdof}, this property allows us
to bound the effect of correlated alignment set selection and obtain an asymptotically accurate sDoF performance bound.
}}
\vspace{3mm}

Based on Lemma~\ref{lem:dim_l}, the following theorem characterizes the sDoF of the GIA algorithm in a stochastic network.

\begin{Thm}[Performance of GIA Algorithm]\label{thm:sDoF}
Define indicator $R\in\mathbb{R}$ as
\begin{eqnarray}
R = \min\Bigg\{1-\frac{\mc{N}{e}}{\mc{\rho}{\ml}\mc{d}{\ml}+\mc{\rho}{j}\mc{d}{j}},\frac{\mc{N}{\ml}\!-\!\mc{d}{\ml}
+}{}\nonumber\hspace{15mm}
\\
\frac{
\min\{\mc{m}{\ml}\mc{d}{\ml},\mc{\rho}{\ml}\mc{d}{\ml}\}\!+\!\min\{\frac{\mc{\rho}{j}}{\mc{\rho}{l}}\mc{m}{j}\mc{d}{j},
\mc{\rho}{j}\mc{d}{j}\}}{\mc{\rho}{\ml}\mc{d}{\ml}+\mc{\rho}{j}\mc{d}{j}}\!-\!1\!\Bigg\},
\label{eqn:R}
\end{eqnarray}

When $\mc{\rho}{\ml}\ge 1$ and $|R|>\sqrt{\frac{\max\left\{\mc{d}{\ml},\mc{d}{j}\right\}
\max\left\{\mc{\rho}{\ml},\mc{\rho}{j}\right\}
}{\big(\mc{\rho}{\ml} \big)^{2}\mc{d}{\ml}}}
$, the sDoF per node  $D^{\mathrm{S}}\in[0,\mc{d}{\ml}]$ is given by
\begin{eqnarray}
\mc{d}{\ml}\!\bigg(\mathbb{I}\{R> 0\} + \mathcal{O}\Big(\frac{\max\left\{\mc{d}{\ml},\mc{d}{j}\right\}\max\left\{\mc{\rho}{\ml},\mc{\rho}{j}\right\}}
{\big(\mc{\rho}{\ml} R\big)^{2}\mc{d}{\ml}}
\Big)\!\bigg),\label{eqn:sdof_T}
\end{eqnarray}
where $\mc{m}{\ml}$, $\mc{m}{j}$ are defined in \eqref{eqn:ml}.
\end{Thm}

\begin{IEEEproof}  Please refer to Appendix~\ref{pf_thm:sDoF} for the proof.
\end{IEEEproof}

\begin{figure}[t] \centering
\includegraphics[scale=0.85]{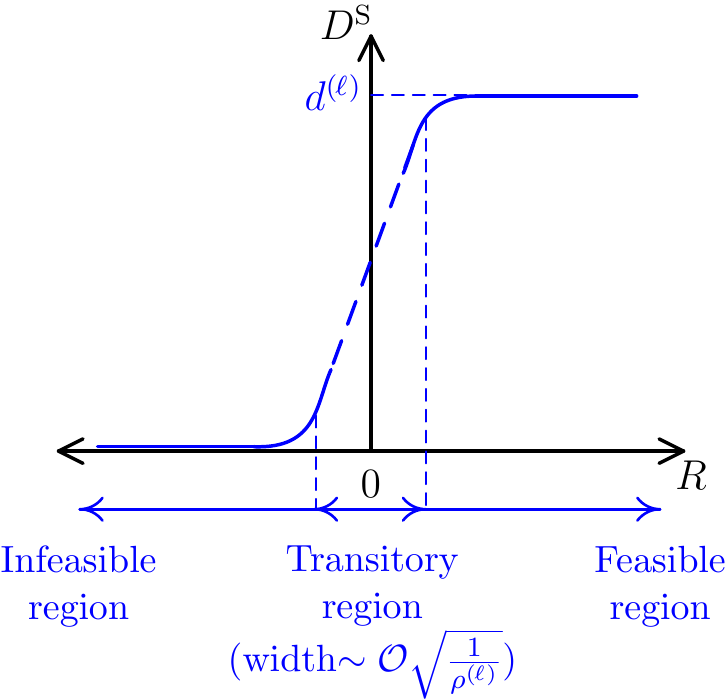}
\caption {Illustration of the sDoF per node described by \eqref{eqn:sdof_T}.} \label{fig_sdof}\vspace{-3mm}
\end{figure}

\begin{Remark}[Interpretation of Theorem~\ref{thm:sDoF}]
\label{remark:sdof}Fig.~\ref{fig_sdof} gives an intuitive illustration of the meaning of the sDoF expression in \eqref{eqn:sdof_T}.
This expression partitions the operation region into three parts according to the value of the indicator $R$.
The sDoF per LR $D^{\mathrm{S}}$ is close to the upper bound $\mc{d}{\ml}$ in the feasible region, whereas it is close to the lower bound $0$ in the infeasible region.
Since the width of the transitory region is on $\mathcal{O}\sqrt{\frac{1}{\mc{\rho}{\ml}}}$, \eqref{eqn:sdof_T} is asymptotically accurate when $\mc{\rho}{\ml}\rightarrow \infty$.
This trend will be shown via simulation in Fig.~\ref{fig_step}.
~\hfill~\IEEEQED
\end{Remark}

In practice, it is interesting to understand how a stochastic wireless-tap network performs under various network parameters.
However, in general, as data stream numbers $\mc{d}{\ml}$, $\mc{d}{j}$ must be integers, and \eqref{eqn:R} contains the discontinuous function $\lfloor\cdot\rfloor$, it is difficult to obtain simple insights.
To address this issue, a network with high connection density, i.e., $\mc{\rho}{\ml}\gg 1$, is analyzed in Appendix~\ref{sec:dense}.
In this scenario, the width of the transitory region is ignorable, and hence sDoF per node $D^{\mathrm{S}}=\mc{d}{\ml}$ when indicator $R>0$. Define the set of feasible data streams as  $\{(\mc{d}{j},\mc{d}{\ml}):R>0\}$.
Analysis on this feasible set reveals insights into the role of GIA in secrecy enhancement, as well as the effect of the network parameters on the performance of wireless-tap networks.
These insights are summarized in the following two remarks.

\begin{figure}[t] \centering
\includegraphics[scale=0.9]{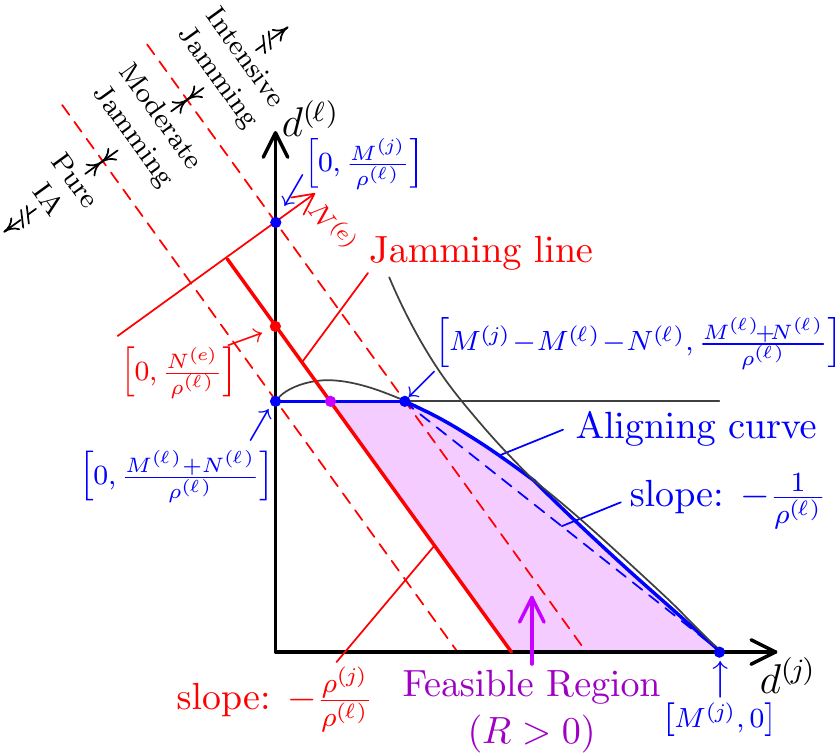}
\caption {Analysis of the feasible region of a stochastic wireless-tap network with high connection density. In this figure, $\mc{M}{j}\ge\mc{M}{\ml}+\mc{N}{\ml}$.}  \label{fig_jam}
\end{figure}

\begin{Remark}[Operation Modes of GIA]
\label{remark:mode}
As illustrated in Fig.~\ref{fig_jam}, in a wireless-tap network with high connection density, the set of feasible streams is contained by the region above the jamming line and below the aligning curve.
The jamming line means that LTs and LJs have generated just enough interference to occupy the signal space of the ERs, and the aligning curve indicates that the LTs, LJs, and LRs are on the cutting edge of being able to align all interference at the LRs.
The slope and the intersection of the jamming line are $-\frac{\mc{\rho}{j}}{\mc{\rho}{\ml}}$ and $\frac{\mc{N}{e}}{\mc{\rho}{\ml}}$, respectively.
The aligning curve is a combination of a horizontal line and two second order curves.
In particular, when $\mc{M}{j}\ge \mc{M}{\ml} + \mc{N}{\ml}$, the trapezoid with vertices $(0,0)$, $(0,\frac{\mc{M}{\ml}+\mc{N}{\ml}}{\mc{\rho}{\ml}})$, $(\mc{M}{j}-\mc{M}{\ml}-\mc{N}{\ml},\frac{\mc{M}{\ml}+\mc{N}{\ml}}{\mc{\rho}{\ml}})$ and $(\mc{M}{j},0)$ lies below the aligning curve.
From Fig.~\ref{fig_jam}, GIA has three operation modes:
\begin{itemize}
\item{\bf Pure IA mode:} When $\mc{N}{e}\le \mc{N}{\ml}+\mc{M}{\ml}$, the LTs and LRs can generate sufficient interference to jam the ERs and align all interference at the LRs. The LJs can remain idle without losing optimality in the sDoF sense.
\item{\bf Moderate Jamming mode:} When  $\mc{N}{\ml}+\mc{M}{\ml}<\mc{N}{e}\le \max\left\{\mc{M}{j},\mc{N}{\ml}+\mc{M}{\ml}\right\}$, by adopting a small $\mc{d}{j}$, the LJs can help the LTs to jam the ERs without reducing the sDoF per LR.
\item{\bf Intensive Jamming mode:} When  $\mc{N}{e}>\max\left\{\mc{M}{j},\mc{N}{\ml}+\mc{M}{\ml}\right\}$, the LJs need to adopt a large $\mc{d}{j}$ to generate sufficient interference to jam the ERs. As a cost of large $\mc{d}{j}$, the sDoF per LR $\mc{d}{\ml}$ needs to be reduced so as to align the interference at the LRs.~\hfill~\IEEEQED
\end{itemize}
\end{Remark}

\begin{figure}[t] \centering
\includegraphics[scale=0.75]{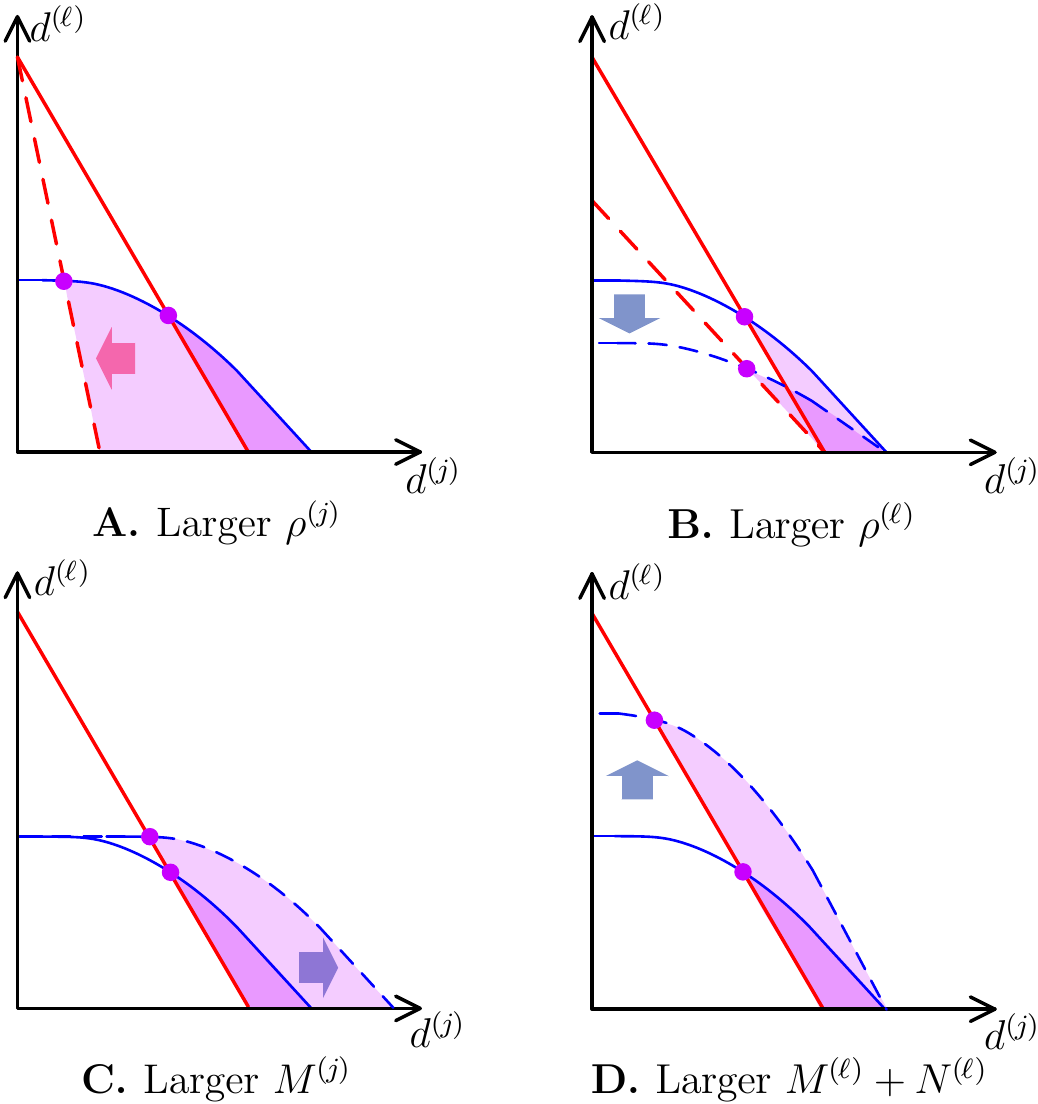}
\caption {Illustration of how the feasible region of $(\mc{d}{j},\mc{d}{\ml})$ changes w.r.t. to network parameters. } \label{fig_parameter}
\end{figure}

\begin{Remark}[Role of Network Parameters]
\label{remark:role}
The effects of network parameters on the sDoF performance of a wireless-tap network are summarized below.
\begin{itemize}
\item {\bf LJ density $\mc{\rho}{j}$:} As Fig.~\ref{fig_parameter}A shows, larger $\mc{\rho}{j}$ leads to a steeper jamming line. This will increase achievable sDoF per LR if GIA is in the intensive jamming mode.
\item {\bf LT/LR density $\mc{\rho}{\ml}$:} As Fig.~\ref{fig_parameter}B shows, larger $\mc{\rho}{\ml}$ flattens both the jamming line and aligning curve, which  reduces achievable sDoF per LR.
\item {\bf LJ antenna $\mc{M}{j}$: } As Fig.~\ref{fig_parameter}C shows, larger $\mc{M}{j}$ pushes the aligning line to the right. This will benefit achievable sDoF per LR if GIA is in the intensive jamming mode.
\item {\bf Sum of LT and LR antenna $\mc{M}{\ml}\!+\!\mc{N}{\ml}$:} As Fig.~\ref{fig_parameter}D shows, larger $\mc{M}{\ml}\!+\!\mc{N}{\ml}$ pushes up the aligning curve, and hence increases achievable sDoF per LR.~\hfill~\IEEEQED
\end{itemize}
\end{Remark}

\section{Simulation Results}
In this section, we will perform three numerical tests.
\subsection{Secrecy Rate under Different Strategies}\label{sec:throughput}
\begin{figure}[t] \centering
\includegraphics[scale=0.62]{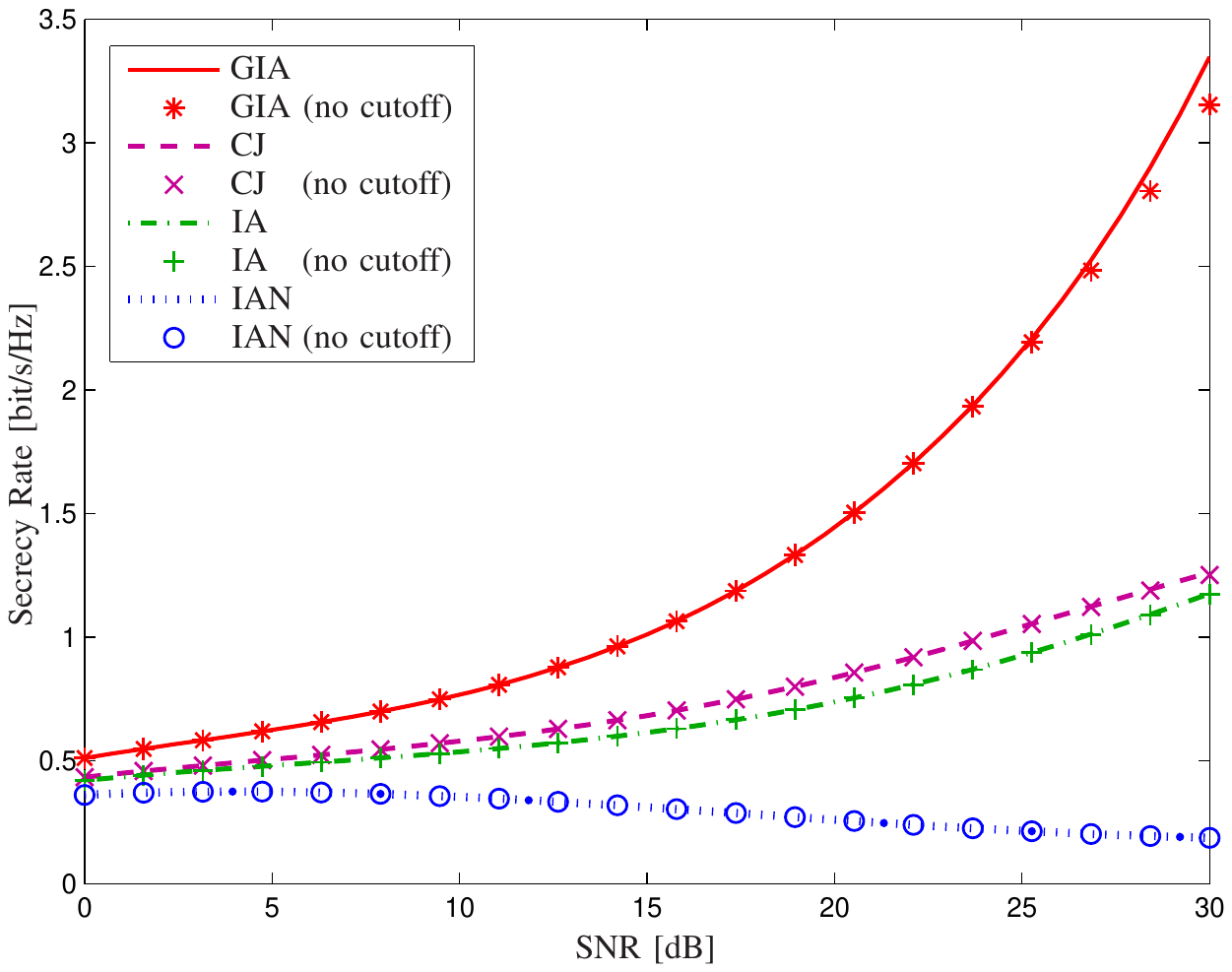}
\caption{Secrecy rate as a function of SNR under different schemes. The network parameters are given by $\alpha=4$, $\theta = 10^{-2}$, $\mc{\lambda}{\ml}=4\times10^{-2}$, $\mc{\lambda}{j}=9\times10^{-2}$, $\mc{M}{\ml}=\mc{M}{j}=16$, $\mc{N}{\ml}=8$, $\mc{N}{e}=32$, and $\mc{d}{\ml}=\mc{d}{j}=1$. The distance between an LT and the associated LR and ER are given by $||\mc{\Delta\mathbf{a}}{\ml}||=1$ and $||\mc{\Delta\mathbf{a}}{e}||=1.5$, respectively.}
\label{fig_throughput}
\end{figure}
First compare the secrecy rate (defined in \eqref{eqn:MIMOC}) achieved by the proposed GIA technique with the following three baselines.
\begin{itemize}
\item{\bf Cooperative jamming (CJ):} The LTs and LRs adopt random transceivers, and the LJs adopt zero-forcing (ZF) precoders to cancel their interference with the LRs.
\item{\bf Pure IA (IA):} The LTs and LRs adopt IA to cancel interference. The LJs remain idle.
\item{\bf IA with artificial noise (IAN):} The LTs and LRs adopt IA to cancel interference. The LJs generate artificial noise by adopting random precoders.
\end{itemize}

To verify the legitimacy of the pathloss model proposed in \eqref{eqn:pathloss}, the secrecy rates under channel models with and without pathloss cutoff are simulated.

Fig.~\ref{fig_throughput} illustrates that the proposed GIA technique achieves significant performance gain over the baselines.
This is because GIA fully exploits the capability of all legitimate partners to create different amounts of interference at the LRs and ERs.
In particular, the low secrecy rate achieved by the IAN scheme highlights the challenge of the concurrent effect of interference.\footnote{The secrecy rate under the IAN scheme slightly drops when the SNR increases. This is because the distance from an LT to the associated LR is smaller than that to the associated ER, which gives the LR an SNR advantage. However, when the SNR increases, this advantage is soon jeopardized by the strong interference from the LJs, which affects both the LRs and ERs.}
From the slope of the secrecy rate under GIA technique, it can be seen that the sDoF per node is around $0.9$. It is not exactly $1$ due to the uncertainty term in \eqref{eqn:sdof_T}.
Also, it can be seen that the secrecy rate under the two types of channel models are reasonably close.

\subsection{Width of the Transitory Region}
\begin{figure}[t] \centering
\psfrag{R}{$\rho$}
\includegraphics[scale=0.62]{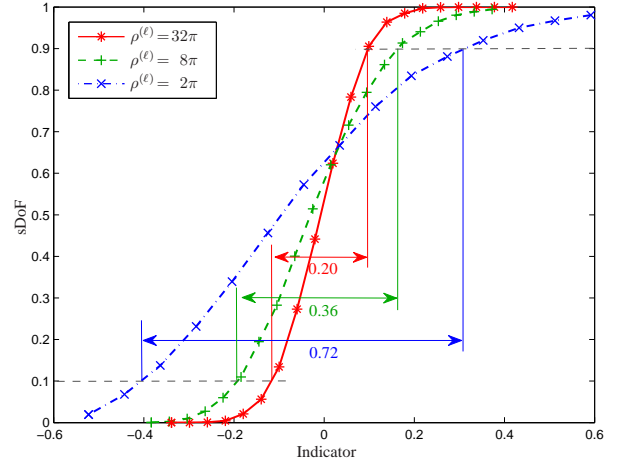}
\caption{sDoF per node $D^{\mathrm{S}}$ as a function of the performance indicator $R$ under different user densities. The network parameters are given by $\alpha=4$, $\theta = 10^{-2}$, and $\mc{d}{\ml}=\mc{d}{j}=1$. The node density is given by
$\mc{\lambda}{\ml}=\mc{\lambda}{j}=32\times10^{-2}$, $\mc{\lambda}{\ml}=\mc{\lambda}{j}=8\times10^{-2}$, and $\mc{\lambda}{\ml}=\mc{\lambda}{j}=2\times10^{-2}$ for the three curves, respectively. Then fix $\mc{M}{\ml}$, $\mc{M}{e}$, $\mc{N}{e}$ and modify $\mc{N}{\ml}$ to change the indicator $R$.}
\label{fig_step}
\end{figure}
Fig.~\ref{fig_step} illustrates the sDoF per node $D^{\mathrm{S}}$ as a function of the indicator $R$ under different network densities. If the region in which $D^{\mathrm{S}}\in[0.1,0.9]\cdot\mc{d}{\ml}$ is used to represent the transitory region, one can see that the width of this region scales on $\mathcal{O}\sqrt{\frac{1}{\mc{\rho}{\ml}}}$. This fact fits the trend described in Remark~\ref{remark:sdof}.

\subsection{Resource Allocation between Transmitting and Jamming}
\begin{figure}[t] \centering
\includegraphics[scale=0.81]{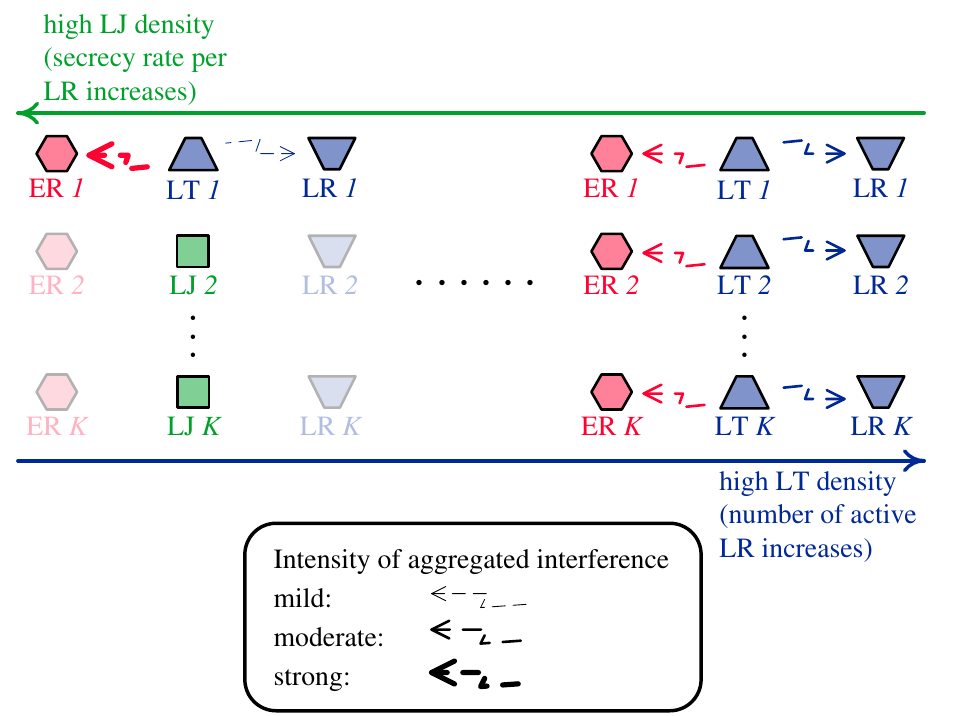}
\caption{Conceptual illustration of the effect of the resource allocation between transmitting and jamming.}
\label{fig_convert}
\end{figure}

So far,  LTs and LJs are assumed to have a fixed prior role. However, their respective roles may overlap:
As illustrated in Fig.~\ref{fig_convert}, in a wireless-tap network, if part of the LRs are deactivated, then from the point of view of the remaining network nodes, the corresponding LTs effectively become LJs.
This conversion empowers the possibility of allocating resources between transmitting and jamming.
The comparison between the left and right column of Fig.~\ref{fig_convert} sketches the effect of allocating resources between transmitting and jamming: one can increase the secrecy rate of each LR by having more LTs deliver dummy information or vice versa.
This effect can also be interpreted from Fig.~\ref{fig_parameter}. The operation of turning LTs to LJs is  equivalent to increase $\mc{\rho}{j}$ and decrease $\mc{\rho}{\ml}$. From Fig.~\ref{fig_parameter}A, B, this operation enlarges the feasible region and hence increases the sDoF per node at a cost of having less active LRs.

\begin{figure*}[t] \centering\hspace{-3mm}
\includegraphics[scale=0.85]{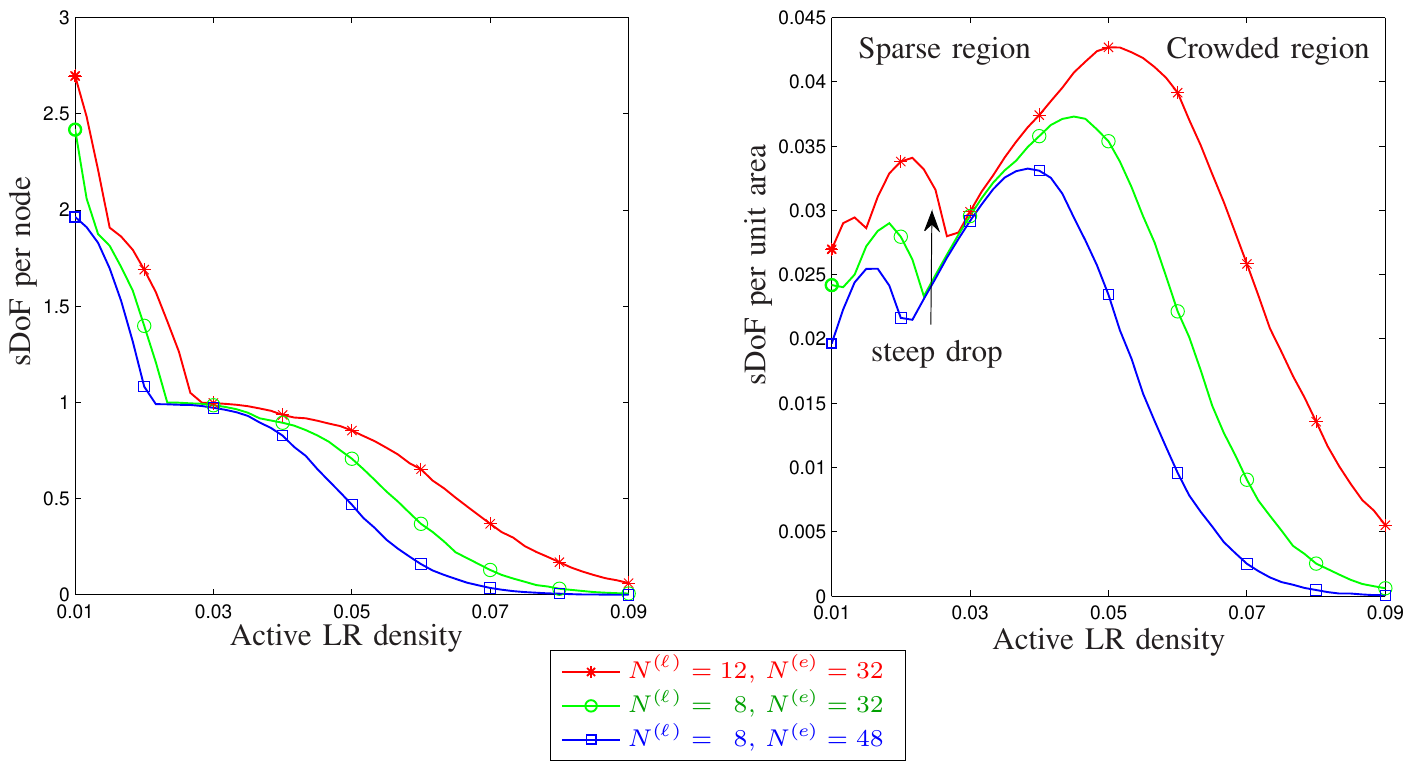}
\caption{sDoF per node/unit area as a function of active LR density $\mc{\lambda}{\ml}$. The network parameters are given by $\alpha=4$, $\theta = 10^{-2}$, $\mc{\lambda}{\ml}\in[1\times10^{-2},9\times10^{-2}]$, $\mc{\lambda}{j}=13\times10^{-2}-\mc{\lambda}{\ml}$, and $\mc{M}{\ml}=\mc{M}{j}=16$.}
\label{fig_tradeoff}
\end{figure*}

Fig.~\ref{fig_tradeoff} illustrates the effect of resource allocation between transmitting and jamming.
We fix the sum of the density of the LTs and LJs, i.e., $\mc{\lambda}{\ml}+\mc{\lambda}{j}$, and illustrate the sDoF per node $D^{\mathrm{S}}$, or per unit area $D^{\mathrm{S}}\mc{\lambda}{\ml}$ as functions of the density of LTs $\mc{\lambda}{\ml}$ (note that this is also the density of active LRs).
Under each active LR density, all the possible stream combinations $(\mc{d}{j},\mc{d}{\ml})$ are exhaustively searched to pick out the combination that gives the highest $D^{\mathrm{S}}$.
From the left column of Fig.~\ref{fig_tradeoff}, it can be observed that sDoF per node $D^{\mathrm{S}}$ is higher, with larger $\mc{N}{\ml}$ and smaller $\mc{N}{e}$ or $\mc{\lambda}{\ml}$. These trends are consistent with those indicated in Remark~\ref{remark:mode} and~\ref{remark:role}.
From the right column of Fig.~\ref{fig_tradeoff}, in terms of sDoF per unit area, one can roughly separate the operation region into two parts, namely the sparse region and the crowded region.
In the sparse region, the benefit from more active LRs dominates, and hence the sDoF per unit area $D^{\mathrm{S}}\mc{\lambda}{\ml}$ increases under larger $\mc{\lambda}{\ml}$.\footnote{
This fact is not always true due to the discrete choices of $(\mc{d}{j},\mc{d}{\ml})$.
For instance, the steep drop indicated in the figure occurs when $\mc{d}{\ml}$ changes from 2 to 1.
The discrete choices of $(\mc{d}{j},\mc{d}{\ml})$ make the sDoF lines rough in some operation regions.}
In the crowded region, the loss from smaller sDoF per node $D^{\mathrm{S}}$ dominates, and hence $D^{\mathrm{S}}\mc{\lambda}{\ml}$ becomes a decreasing function of $\mc{\lambda}{\ml}$.
Therefore, in practice, it is important to control active LR density so that the network operates in a favorable region.

\section{Summary}
\label{sec:conclude}
By creating strong interference at the ERs  but little or no interference at the LRs, the GIA technique provides an effective tool for wireless secrecy protection.
Based on the theoretical framework established in Part I, in this part, we designed a GIA algorithm that is applicable to large-scale networks and characterized the performance of this algorithm in stochastic wireless-tap networks.
We have identified the working modes of GIA and obtained simple insights into how network parameters affect the performance of wireless-tap networks that adopt the GIA technique.
Numerical results illustrate the contribution of GIA in wireless secrecy protection and confirms the obtained insights.

\appendices
\section{Proof of Theorem~\ref{thm:cover}}
\label{pf_thm:cover}
From Theorem~4.4 of Part I, one only need to show that matrix $\mathbf{H}_{\mathrm{all}}$ (defined in Fig.~4 of Part I) is full row-rank.
Suppose Problem~\ref{pro:GIA} is feasible under alignment subsets $\{\mc{\mathcal{A}}{r}(k)\}$ and $\{\mc{\mathcal{A}}{t}(j)\}$.
If the intersection of some alignment subsets are non-empty, e.g., $\mc{\mathcal{A}}{r}(k)\cap\mc{\mathcal{A}}{t}(j)\neq \emptyset$,
then non-overlapping alignment subset $\tilde{\mc{\mathcal{A}}{r}}(k)=\mc{\mathcal{A}}{r}(k)\backslash (\mc{\mathcal{A}}{r}(k)\cap\mc{\mathcal{A}}{t}(j))$ can be generated. From Corollary~4.1 of Part I, since this operation does not change the alignment set $\mathcal{A}$, the feasibility of Problem~\ref{pro:GIA} is preserved. Hence, to prove the theorem, it is sufficient to consider the case in which
\begin{eqnarray}
\mc{\mathcal{A}}{r}(k)\cap\mc{\mathcal{A}}{t}(j)= \emptyset, \qquad \forall k,j.\label{eqn:nooverlap}
\end{eqnarray}

From equations \eqref{eqn:cover} and \eqref{eqn:nooverlap}, every $(k,j)\in\mathcal{A}$ belongs to one and only one alignment subset. Hence, one can reorder the rows of $\mathbf{H}_{\mathrm{all}}$ and rewrite the matrix as
\begin{eqnarray}
\mathrm{diag}(\mathbf{H}^{\mathrm{U}}_1,\ldots,\mathbf{H}^{\mathrm{U}}_K,
\mathbf{H}^{\mathrm{V}}_1,\ldots,\mathbf{H}^{\mathrm{V}}_{\tilde{K}}) + \tilde{\mathbf{H}}_{\mathrm{all}},\label{eqn:H_all_reorder}
\end{eqnarray}
where
\begin{eqnarray}
\mathbf{H}^{\mathrm{U}}_k=
\begin{bmatrix}\mathbf{H}^{\mathrm{U}}_{kj_1}
\\\vdots\\
\mathbf{H}^{\mathrm{U}}_{kj_S}
\end{bmatrix}, \qquad (k,j_s)\in \mc{\mathcal{A}}{r}(k),\; S = |\mc{\mathcal{A}}{r}(k)|, \label{eqn:huk}
\end{eqnarray}
\begin{eqnarray}
\mathbf{H}^{\mathrm{V}}_j=
\begin{bmatrix}\mathbf{H}^{\mathrm{V}}_{k_1j}
\\\vdots\\
\mathbf{H}^{\mathrm{V}}_{k_Sj}
\end{bmatrix}, \qquad (k_s,j)\in \mc{\mathcal{A}}{t}(j),\; S = |\mc{\mathcal{A}}{t}(j)|, \label{eqn:hvj}
\end{eqnarray}
and the submatrices in $\tilde{\mathbf{H}}_{\mathrm{all}}$ are given by
\begin{eqnarray}
\mathbf{H}^{\mathrm{U}}_{kj}:\; (k,j)\not\in \mc{\mathcal{A}}{r}(k),\quad\mbox{and}\quad
\mathbf{H}^{\mathrm{V}}_{kj}:\; (k,j)\not\in \mc{\mathcal{A}}{t}(j).\label{eqn:notH}
\end{eqnarray}
Substituting the condition of proper alignment subset, i.e., \eqref{eqn:proper}, to the expressions of $\mathbf{H}^{\mathrm{U}}_{kj}$ and $\mathbf{H}^{\mathrm{V}}_{kj}$, i.e., equations (9) and (10) of Part I, we get that matrices $\mathbf{H}^{\mathrm{U}}_k$ and $\mathbf{H}^{\mathrm{V}}_j$ in \eqref{eqn:huk} and \eqref{eqn:hvj} are full row-rank almost surely. Hence, $\mathrm{diag}(\mathbf{H}^{\mathrm{U}}_1,\ldots,\mathbf{H}^{\mathrm{V}}_{\tilde{K}})$ is full row-rank almost surely.
Moreover, from equations (9) and (10) of Part I, the elements in different submatrices $\mathbf{H}^{\mathrm{U}}_{kj}$, $\mathbf{H}^{\mathrm{V}}_{kj}$ are independent. Hence \eqref{eqn:notH} assures that $\tilde{\mathbf{H}}_{\mathrm{all}}$ is independent of $\mathrm{diag}(\mathbf{H}^{\mathrm{U}}_1,\ldots,\mathbf{H}^{\mathrm{V}}_{\tilde{K}})$. Therefore, from \eqref{eqn:H_all_reorder}, $\mathbf{H}_{\mathrm{all}}$ is full row-rank almost surely. This completes the proof.

\section{Proof of Lemma~\ref{lem:SDI}}
\label{pf_lem:SDI}
As the entries of the channel matrices are independent random variables drawn
from continuous distributions, with probability 1,
\begin{align}
\mc{r}{\iota}=\mc{S}{\iota}\log_2(P)+\scalebox{0.8}{$\mathcal{O}$}(\log_2
P)
\end{align}
Substituting this result to \eqref{eqn:MIMOC},
\begin{align}
R^{\mathrm{S}}=\big[\mc{S}{\ml}-\mc{S}{e}\big]^+\log_2(P)
+\scalebox{0.8}{$\mathcal{O}$}(\log_2P)\label{eqn:RS2}
\end{align}
Substituting \eqref{eqn:RS2} to \eqref{eqn:sDoF},  \eqref{eqn:sDoF2} is obtained.
In the following, the expression of $\mc{S}{\ml}$ and $\mc{S}{e}$ will be derived.

If a link between LR $\mathbf{a}$ and LT (or LJ) $\mathbf{b}$
has zero pathloss, i.e., $L(\mathbf{a},\mathbf{b})=0$, or
$(\mathbf{a},\mathbf{b})\in\mathcal{A}$, there is no interference
on this link.
Otherwise, the channel state $\MNC{H}{\ml}{\mathbf{a},\mathbf{b}}$ is independent
of $\MNC{U}{\ml}{\mathbf{a}}, \mathbf{V}_{\mathbf{b}}$.
In this case, $\mathrm{rank}\Big(\big(\MNC{U}{\ml}{\mathbf{a}}\big)^\mathrm{H}
\mathbf{H}_{\mathbf{a},\mathbf{b}}\mathbf{V}_{\mathbf{b}}\Big)=\min\{\mc{d}{\ml},\mc{d}{x}\}$
almost surely, where $x=\ml$ for LTs and $x=j$ for LJs.
Hence, with
probability 1, $\mc{S}{\ml}$ is given by \eqref{eqn:Sl}.

Similarly, as the channel state of the eavesdropping network $\{\MNC{H}{e}{\mathbf{e}\mathbf{b}}\}$
is independent of precoders $\{\mathbf{V}_{\mathbf{b}}\}$, \eqref{eqn:Se} is obtained.

\section{Proof of Lemma~\ref{lem:dim_l}}
\label{pf_lem:dim_l}
First prove \eqref{eqn:Sl2}.
$\mathcal{A}=\left(\cup_{\mathbf{a}\in \mc{\mathcal{R}}{\ml}}
\mc{\mathcal{A}}{r}(\mathbf{a})\right)\cup
\left(\cup_{\mathbf{b}\in \mc{\mathcal{T}}{\ml}}\mc{\mathcal{A}}{t}(\mathbf{b})\right)
\cup \left(\cup_{\mathbf{b}\in \mc{\mathcal{T}}{j}}\mc{\mathcal{A}}{j}(\mathbf{b})\right)$.
From \eqref{eqn:setnooverlap}, the sets $\cup_{\mathbf{a}\in \mc{\mathcal{R}}{\ml}} \mc{\mathcal{A}}{r}(\mathbf{a})$,
$
\cup_{\mathbf{b}\in \mc{\mathcal{T}}{\ml}}\mc{\mathcal{A}}{t}(\mathbf{b})$,
and
$\cup_{\mathbf{b}\in \mc{\mathcal{T}}{j}}\mc{\mathcal{A}}{j}(\mathbf{b})$
do not overlap. Hence,
\begin{eqnarray}
&&\hspace{-10mm}\sum_{{\mathbf{b}\in\mc{\mathcal{T}}{x}}
}\hspace{-1mm} \mathbb{I}\{L(\mathbf{a},\mathbf{b})>0\;\&\;(\mathbf{a},\mathbf{b})\not\in
\mathcal{A}\}=\nonumber\\
&& \hspace{-7mm}\sum_{{\mathbf{b}\in\mc{\mathcal{T}}{x}}\atop (\mathbf{a},\mathbf{b})\not\in
\mc{\mathcal{A}}{x}(\mathbf{b})}\hspace{-3mm} \mathbb{I}\{L(\mathbf{a},\mathbf{b})>0\}-
\hspace{-3mm}\sum_{{\mathbf{b}\in\mc{\mathcal{T}}{x}}\atop (\mathbf{a},\mathbf{b})\in
\mc{\mathcal{A}}{r}(\mathbf{a})}\hspace{-3mm}\mathbb{I}\{L(\mathbf{a},\mathbf{b})>0\}\label{eqn:Ix},
\end{eqnarray}
where $x\in\{\ml,j\}$. Substituting \eqref{eqn:Il} and \eqref{eqn:Ix} to \eqref{eqn:Sl},
\begin{align}
\mc{S}{\ml}=\big[\mc{N}{\ml}-\mc{I}{\ml}-\tilde{\varepsilon}\big]^+,\label{eqn:Sl3}
\end{align}
where $\tilde{\varepsilon}=\mc{d}{\ml}\sum_{{\mathbf{b}\in\mc{\mathcal{T}}{\ml}}\atop (\mathbf{a},\mathbf{b})\in
\mc{\mathcal{A}}{r}(\mathbf{a})}\mathbb{I}\{L(\mathbf{a},\mathbf{b})>0\}
+\mc{d}{j}\sum_{{\mathbf{b}\in\mc{\mathcal{T}}{j}}\atop
(\mathbf{a},\mathbf{b})\in
\mc{\mathcal{A}}{r}(\mathbf{a})}\mathbb{I}\{L(\mathbf{a},\mathbf{b})>0\}$.
From \eqref{eqn:epsilon}, $\tilde{\varepsilon}\le\varepsilon$. Moreover, from \eqref{eqn:nearR} and \eqref{eqn:nearT}, if $\tilde{\varepsilon}<\varepsilon$,  $\big\{\mathbf{b}\in\mc{\mathcal{T}}{\ml}\cup\mc{\mathcal{T}}{j}\backslash\{\mathbf{a}+\mathbf{p}_\mathbf{a}\}
:L(\mathbf{a},\mathbf{b})>0\big\}\subseteq\mathcal{A}$, which means $\mc{S}{\ml}=\mc{d}{\ml}$. With this fact and \eqref{eqn:Sl3},\eqref{eqn:Sl2} is obtained.

From \eqref{eqn:Set_r}, it is easy to see that  $\varepsilon$ is bounded within $\big[0,\max\{\mc{d}{j},\mc{d}{\ml}\}-1\big]$. Hence, in the following, the focus is on characterizing the mean and variance of $\mc{I}{\ml}$.
Define
$
\mc{i}{x} \triangleq \sum_{{\mathbf{b}\in\mc{\mathcal{T}}{x}}\atop (\mathbf{a},\mathbf{b})\not\in
\mc{\mathcal{A}}{x}(\mathbf{b})}\mathbb{I}\{L(\mathbf{a},\mathbf{b})>0\}$,
$x\in\{\ml,j\}$, then
\begin{align}
\mc{I}{\ml}=\mc{d}{\ml}\mc{i}{\ml}+\mc{d}{j}\mc{i}{j}.\label{eqn:Il2}
\end{align}

To analyze the mean and variance of  $\mc{I}{\ml}$, we first analyze those of $
\mc{i}{x}$, $x\in\{\ml,j\}$.
To achieve this task,  a characterization  the spatial distribution of LRs is needed.
\begin{Lem}[Spatial Distribution of LRs]\label{lem:LRdis} In a stochastic
network, as described by Definition~\ref{def:typology}, the position
of the LRs is given by a PPP with density $\mc{\lambda}{\ml}$.
\end{Lem}
\begin{IEEEproof} From the second item in Definition~\ref{def:typology},
the position of the LRs is a transformation of that of the LTs, which
is a PPP with density $\mc{\lambda}{\ml}$. Hence, from \cite[Thm. 1.3.9]{BacBls:B09},
the position of the LRs is also a PPP with density $\lambda(\mathbf{a})$,
$\mathbf{a}\in\mathbb{R}^2$, where
\begin{eqnarray}
\lambda(\mathbf{a})=\int_{\mathbb{R}^2}p\big(\mc{\Delta\mathbf{a}}{\ml}\big)\mc{\lambda}{\ml}
d\big(\mc{\Delta\mathbf{a}}{\ml}\big)=\mc{\lambda}{\ml}.
\end{eqnarray}
Here $p(\mathbf{x})$ denotes the probability density function of $\mathbf{x}$.
This completes the proof.
\end{IEEEproof}

First analyze the expectation $\mc{i}{x}$. For LR $\mathbf{a}$, the positions of the unassociated LTs are given by a homogeneous PPP with density $\mc{\lambda}{\ml}$ on $\mathbb{R}^2\backslash\{\mathbf{a}+\mathbf{p}_{\mathbf{a}}\}$. Hence
\begin{eqnarray}
\hspace{-5mm}\mathbb{E}\left\{\mc{i}{x}\right\}&\skiph=\skiph&\int_{0}^
{\theta^{-2/\alpha}}2\pi r \mc{\lambda}{x}\Pr\Big\{(\mathbf{a},\mathbf{b})\not\in
\mc{\mathcal{A}}{x}\big|\nonumber
\\&&
\mathbf{b}\in\mc{\mathcal{T}}{x}, ||\mathbf{a}-\mathbf{b}||=r
\Big\} \mathrm{d}r.\label{eqn:ix_1}
\end{eqnarray}

From Lemma~\ref{lem:LRdis},
\begin{eqnarray}
&&\hspace{-10mm}\nonumber\Pr\Big\{(\mathbf{a},\mathbf{b})\not\in
\mc{\mathcal{A}}{x}\big|\mathbf{b}\in\mc{\mathcal{T}}{x}, ||\mathbf{a}-\mathbf{b}||=r\Big\}
\\&&\hspace{-10mm}= 1\!-\!\Pr\Big\{\big|\big\{\mathbf{a}'\!:\mathbf{a}'\!\in
\mc{\mathcal{R}}{\ml},||\mathbf{a}'-\mathbf{b}||<r\big\}
\big|\le
\mc{m}{x}\!-\!1\Big\}\label{eqn:Pr_mx}
\\&&\hspace{-10mm}= 1-\frac{\Gamma(\mc{m}{x},\pi r^2 \mc{\lambda}{\ml})}{\Gamma(\mc{m}{x})}.
\label{eqn:Pr_gamma}
\end{eqnarray}

Substitute \eqref{eqn:Pr_gamma} to \eqref{eqn:ix_1}:
\begin{eqnarray}
\nonumber\mathbb{E}\left\{\mc{i}{x}\right\}&\skiph=&\skiph \pi\theta^{-\frac{4}{\alpha}}\mc{\lambda}{x}-\frac{2\pi\mc{\lambda}{x}}{\Gamma(\mc{m}{x})}
\int_{0}^
{\theta^{-2/\alpha}}\hspace{-4mm} r
\Gamma(\mc{m}{x},\pi r^2 \mc{\lambda}{\ml}) \mathrm{d}r
\\&\skiph=&\skiph\mc{\rho}{x}-
\frac{\mc{\lambda}{x}}{\mc{\lambda}{\ml}}\mc{m}{x}-\frac{\mc{\lambda}{x}}{\mc{\lambda}{\ml}}\nonumber
\\&&\hspace{-3mm}
\frac{(\mc{\rho}{\ml}-\mc{m}{x})\Gamma(\mc{m}{x},\mc{\rho}{\ml})
-(\mc{\rho}{\ml})^{\mc{m}{x}}e^{-\mc{\rho}{\ml}}\hspace{-2mm}
}{\Gamma(\mc{m}{x})}. \label{eqn:ix_2}
\end{eqnarray}

From \cite[8.11.2]{MathFun:B09},
\begin{eqnarray}\lim_{\mc{\rho}{\ml}\rightarrow +\infty}\hspace{-3mm}\frac{(\mc{\rho}{\ml}-\mc{m}{x})
\Gamma(\mc{m}{x},\mc{\rho}{\ml})
-(\mc{\rho}{\ml})^{\mc{m}{x}}e^{-\mc{\rho}{\ml}}
}{\Gamma(\mc{m}{x})}=0.\label{eqn:lim}
\end{eqnarray}

By combining \eqref{eqn:Pr_mx}, \eqref{eqn:ix_2}, and \eqref{eqn:lim},
\begin{align}
&\nonumber-\frac{\mc{\lambda}{x}}{\mc{\lambda}{\ml}}
\frac{(\mc{\rho}{\ml}-\mc{m}{x})\Gamma(\mc{m}{x},\mc{\rho}{\ml})
-(\mc{\rho}{\ml})^{\mc{m}{x}}e^{-\mc{\rho}{\ml}}
}{\Gamma(\mc{m}{x})}=\hspace{1.3cm}
\\&\int_{\theta^{-2/\alpha}}^
{+\infty}\hspace{-3mm}2\pi r \mc{\lambda}{x}
\!\Pr\big\{\big|\big\{\mathbf{a}'\!:\mathbf{a}'\!\in\mc{\mathcal{R}}{\ml},||
\mathbf{a}'\!-\!\mathbf{b}||<r\big\}\big|\le
\mc{m}{x}\!-\!1\big\}\nonumber
\\& \mathrm{d}r,\nonumber
\end{align}
which is a positive, increasing function of $\mc{m}{x}$. Hence, when $\mc{m}{x}\le\mc{\rho}{\ml}$,
$\mathbb{E}\left\{\mc{i}{x}\right\}$ is in interval
\begin{eqnarray}
\nonumber &&\hspace{-10mm}\left[\mc{\rho}{x}-
\frac{\mc{\lambda}{x}}{\mc{\lambda}{\ml}}\mc{m}{x} ,\mc{\rho}{x}-
\frac{\mc{\lambda}{x}}{\mc{\lambda}{\ml}}\mc{m}{x}+\frac{\mc{\lambda}{x}}{\mc{\lambda}{\ml}}
\frac{(\mc{\rho}{\ml})^{\mc{\rho}{\ml}}e^{-\mc{\rho}{\ml}}
}{\Gamma(\mc{\rho}{\ml})} \right]
\\& &\hspace{-10mm}\subseteq\left[\mc{\rho}{x}-
\frac{\mc{\lambda}{x}}{\mc{\lambda}{\ml}}\mc{m}{x} ,\mc{\rho}{x}-
\frac{\mc{\lambda}{x}}{\mc{\lambda}{\ml}}\mc{m}{x}+
\frac{\mc{\lambda}{x}\sqrt{\mc{\rho}{\ml}}}{\mc{\lambda}{\ml}
\sqrt{2\pi}} \label{eqn:ix_3}
 \right].
\end{eqnarray}

\eqref{eqn:ix_3} is true because, from Stirling's formula \cite[5.11.7]{MathFun:B09},
 \begin{eqnarray}\sqrt{\frac{2\pi}{z}}(z)
^{z}e^{-z}\le\Gamma(z)\le\sqrt{\frac{2\pi}{z}}(z)
^{z}e^{-z}(1+\frac{1}{12z})\label{eqn:gamma}
\end{eqnarray}

Further noting that $\mathbb{E}\left\{\mc{i}{x}\right\}$ is a nonnegative
decreasing function of $\mc{m}{x}$, we have $\mathbb{E}\left\{\mc{i}{x}\right\}$
is in interval
\begin{eqnarray}
&&\hspace{-15mm}\bigg[\Big[\mc{\rho}{x}-
\frac{\mc{\lambda}{x}}{\mc{\lambda}{\ml}}\mc{m}{x}\Big]^+,\nonumber
\\&&\hspace{-13mm}\Big[\mc{\rho}{x}-
\frac{\mc{\lambda}{x}}{\mc{\lambda}{\ml}}\mc{m}{x}\Big]^++
\frac{\mc{\lambda}{x}\sqrt{\mc{\rho}{\ml}}}{\mc{\lambda}{\ml}
\sqrt{2\pi}}\bigg].\label{eqn:ix_4}
\end{eqnarray}

We then start to bound the variance of $\mc{i}{x}$. To address the challenge of correlated alignment set selection, the following lemma is proposed.

\begin{Lem}[Bound of the Variance of the Sum of Random Variables]\label{lem:variance} $\{x_1,\ldots,x_n\}$ are random variables in $\mathbb{R}$. Then
\begin{eqnarray}\mathbb{S}\left\{\sum_{i=1}^{n}x_i\right\}
\le\sum_{i=1}^{n}\mathbb{S}\{x_i\}.
\end{eqnarray}
\end{Lem}
\begin{IEEEproof} Denote $\bar{x}_i=\mathbb{E}\{x_i\}$, then
\begin{eqnarray}
\hspace{-4mm}\nonumber\mathbb{V}\left\{\sum_{i=1}^{n}x_i\right\}&\skiph=&\skiph
\sum_{i=1}^{n}\sum_{j=1}^{n}\mathbb{E}\{(x_i-\bar{x}_i)(x_j-\bar{x}_j)\}
\\ &\skiph\le&\skiph \sum_{i=1}^{n}\sum_{j=1}^{n}\sqrt{\mathbb{E}\{(x_i-\bar{x}_i)^2\}\mathbb{E}\{(x_j-\bar{x}_j)^2\}}\label{eqn:cauchy}
\\ &\skiph=&\skiph\left(\sum_{i=1}^{n}\mathbb{S}\{x_i\}\right)^2,
\end{eqnarray}
where \eqref{eqn:cauchy} is true due to the Cauchy--Schwarz inequality. This completes the proof.
\end{IEEEproof}

From Lemma~\ref{lem:variance},
\begin{eqnarray}
\nonumber \mathbb{S}\left\{\mc{i}{x}\right\}&\skiph\le\skiph&\int_{0}^
{\theta^{-2/\alpha}}\hspace{-3mm}2\pi r \mc{\lambda}{x}
\mathrm{S}\Big(\Pr\Big\{(\mathbf{a},\mathbf{b})\not\in \mc{\mathcal{A}}{x}\big|
\mathbf{a}\in\mc{\mathcal{R}}{\ml},
\\&&\hspace{24mm}\mathbf{b}\in\mc{\mathcal{T}}{x} , ||\mathbf{a}-\mathbf{b}||=r
\Big\}\Big) \mathrm{d}r\nonumber
\\&\skiph=\skiph&\int_{0}^
{\theta^{-2/\alpha}}\hspace{-3mm}2\pi r \mc{\lambda}{x}
\mathrm{S}\Big(\Pr\Big\{\Big|
\{\mathbf{a}':\mathbf{a}'\in\mc{\mathcal{R}}{\ml},\nonumber
\\&&\hspace{18mm}||\mathbf{a}'-\mathbf{b}||<r\}
\Big|\ge
\mc{m}{x}\Big\}\Big) \mathrm{d}r
\label{eqn:ixv_1},
\end{eqnarray}
where function $\mathrm{S}(p)=\sqrt{p(1-p)}$, $p\in[0,1]$.

Denote
$\Pr(r,\mc{m}{x})=\Pr\Big\{\big|\{\mathbf{a}':\mathbf{a}'
\in\mc{\mathcal{R}}{\ml},||\mathbf{a}'-\mathbf{b}||<r\}
\big|$ $\ge\mc{m}{x}\Big\}$.
If $\mc{m}{x}=0$, $\Pr(r,\mc{m}{x})=1$, $\forall r\ge 0$. Hence $\mathbb{V}\left\{\mc{i}{x}\right\}=0$. Otherwise, when $\mc{m}{x}\ge 1
$, variable $\big|\big\{\mathbf{a}':\mathbf{a}'\in\mc{\mathcal{R}}{\ml},||\mathbf{a}'-\mathbf{b}||<r\big\}
\big|$ follows Poisson distribution with mean $\pi r^2 \mc{\lambda}{\ml}$.
Hence, from  Chernoff inequality,
\begin{eqnarray}
\hspace{-2mm}\left\{\begin{array}{r@{}l@{}l}
\hspace{-2mm}\Pr(r,\mc{m}{x}) &\le
\frac{\dis e^{-\pi r^2 \mc{\lambda}{\ml}}\hspace{-1mm}\big(e\pi r^2 \mc{\lambda}{\ml}\big)^{\mc{m}{x}}\hspace{-2mm}}
{\dis \big(\mc{m}{x})^{\mc{m}{x}}},
 \mbox{ when: }
r\le R=\!\sqrt{\frac{\mc{m}{x}}{\pi\mc{\lambda}{\ml}}},
\\
\hspace{-2mm}\Pr(r,\mc{m}{x}) &\ge
1-\frac{\dis e^{-\pi r^2 \mc{\lambda}{\ml}}\big(e\pi r^2 \mc{\lambda}{\ml}\big)^{\mc{m}{x}}\hspace{-2mm}}{\dis \big(\mc{m}{x})^{\mc{m}{x}}},
 \mbox{ when: }
r\ge R.
\end{array}\right.\hspace{-8mm}\label{eqn:Pr_bound}
\end{eqnarray}

Then $\mathbb{S}\left\{\mc{i}{x}\right\}$ will be bounded by separating the operation region into the following two cases:

{\bf Case 1:} $R\le \theta^{-2/\alpha}$ (i.e., $\mc{m}{x}\le \mc{\rho}{\ml}$). Substitute \eqref{eqn:Pr_bound} to \eqref{eqn:ixv_1}, noting that $\mathrm{S}(p)\le \min\{\sqrt{p},\sqrt{1-p}\}$, we have
\begin{eqnarray}
\nonumber \hspace{-2mm}\mathbb{S}\big\{\mc{i}{x}\big\}\hspace{-1mm}&\skiph\le\skiph&
\hspace{-1mm}2\pi\mc{\lambda}{x} \Bigg(\hspace{-1mm} \int_{0}^{R}\hspace{-2mm}r
\mathrm{S}\big(\Pr(r,\mc{m}{x}) \big)\mathrm{d}r \!+\hspace{-1.5mm}
\int_{R}^{\theta^{-2/\alpha}}\hspace{-7mm}r
\mathrm{S}\big(\Pr(r,\mc{m}{x}) \big)\mathrm{d}r \hspace{-1mm}\Bigg)\hspace{-5mm}
\\\nonumber &\skiph\le\skiph&\hspace{-1mm}
2\pi\mc{\lambda}{x} \int_{0}^{\theta^{-2/\alpha}}
\frac{r e^{\frac{-\pi r^2 \mc{\lambda}{\ml}}{2}}\big(e\pi r^2 \mc{\lambda}{\ml}\big)^{ \frac{\mc{m}{x}}{2}   }}{ \big(\mc{m}{x})^{\frac{\mc{m}{x}}{2}   }}\mathrm{d}r
\\\nonumber &\skiph\le\skiph&\hspace{-1mm}
-2\pi\mc{\lambda}{x} \frac{(2e)^{\frac{\mc{m}{x}}{2}} \Gamma(\frac{\mc{m}{x}}{2}+1,\frac{\pi\mc{\lambda}{\ml}r^2}{2} )}{ \pi\mc{\lambda}{\ml}\big(\mc{m}{x})^{\frac{\mc{m}{x}}{2}}}\bigg|_{0}^{+\infty}
\\&\skiph\le\skiph&\hspace{-1mm} \frac{4\mc{\lambda}{x}\sqrt{\pi\mc{m}{x}}}{\mc{\lambda}{l}}(1+\frac{1}{6\mc{m}{x}}),
\label{eqn:ixv_2}
\end{eqnarray}
where last inequality is true because of \eqref{eqn:gamma}.

{\bf Case 2:} $R> \theta^{-2/\alpha}$ (i.e., $\mc{m}{x}> \mc{\rho}{\ml}$). First, prove the following lemma.\vspace{1mm}
\begin{Lem}\label{lem:J1}
When $r\in \left[0,\tilde{R}\right]$, $\Pr(r,\mc{m}{x})\le\frac{1}{2}$, where $\tilde{R}=\sqrt{\dis \frac{\mc{m}{x}-\frac{1}{3}}{\dis \pi\mc{\lambda}{\ml}}}$.
\end{Lem}
\begin{IEEEproof} Since $\big|\big\{\mathbf{a}':\mathbf{a}'\in\mc{\mathcal{R}}{\ml},||\mathbf{a}'-\mathbf{b}||<r\big\}
\big|$ is Poisson random variable with mean $\pi r^2 \mc{\lambda}{\ml}$, from \cite[Thm. 2]{Choi:94}, when $\pi r^2 \mc{\lambda}{\ml}+\frac{1}{3}\le \mc{m}{x}$,   $\Pr(r,\mc{m}{x})\le\frac{1}{2}$. This completes the proof.\vspace{2mm}
\end{IEEEproof}

\begin{Lem}\label{lem:J2}
When $r\in \left(\tilde{R},R\right]$ and $\mc{\rho}{\ml}\ge \frac{2}{3}$
$f(r,\mc{\rho}{\ml}) \triangleq \frac{\dis e^{-\pi r^2 \mc{\lambda}{\ml}}\big(e\pi r^2 \mc{\lambda}{\ml}\big)^{\mc{\rho}{\ml}}}{\dis \big(\mc{\rho}{\ml})^{\mc{\rho}{\ml}}}>\frac{1}{2}$.
\end{Lem}
\begin{IEEEproof} Note that when $r\le R$,
\begin{eqnarray}
\frac{\partial f}{\partial r}= \frac{\dis e^{-\pi r^2 \mc{\lambda}{\ml}}\big(e\pi r^2 \mc{\lambda}{\ml}\big)^{\mc{\rho}{\ml}}}{\dis \big(\mc{\rho}{\ml})^{\mc{\rho}{\ml}}}(-2\pi r\mc{\lambda}{\ml} + \frac{2\mc{\rho}{\ml}}{r})\ge 0,
\end{eqnarray}
it can be seen that when $r\in \left(\tilde{R},R\right]$,
\begin{eqnarray}
f(r,\mc{\rho}{\ml})> f(\tilde{R},\mc{\rho}{\ml})  > e^{\mc{\rho}{\ml} \ln (1-\frac{1}{3\mc{\rho}{\ml}})}.\label{eqn:f_rm}
\end{eqnarray}
Noting that $\mc{\rho}{\ml}\ge \frac{2}{3}$,
\begin{eqnarray}
\hspace{-7mm}\nonumber \frac{\partial \mc{\rho}{\ml} \ln (1-\frac{1}{3\mc{\rho}{\ml}})}{\partial \mc{\rho}{\ml}}
&\skiph=\skiph& \ln (1-\frac{1}{3\mc{\rho}{\ml}}) + \frac{1}{3\mc{\rho}{\ml}-1}\\
&\skiph>\skiph& -\sum_{n=1}^{+\infty} \left(\frac{1}{3\mc{\rho}{\ml}}\right)^n + \frac{1}{3\mc{\rho}{\ml}-1}=0\label{eqn:f_rm_partial}
\end{eqnarray}
Substitute \eqref{eqn:f_rm_partial} to \eqref{eqn:f_rm}, we have $f(r,\mc{\rho}{\ml})>f(\tilde{R},\frac{2}{3})>(1-\frac{1}{2})^\frac{2}{3}>\frac{1}{2}$. This completes the proof.
\end{IEEEproof}
With the two lemmas proved above, it can be seen that
\begin{eqnarray}
&&\hspace{-7mm}
\nonumber \mathbb{S}\big\{\mc{i}{x}\big\}
\\\nonumber&\skiph\le\skiph&
2\pi\mc{\lambda}{x} \Bigg(\! \int_{0}^{\min\{\tilde{R},\theta^{-2/\alpha}\}}\hspace{-17mm}r
\mathrm{S}\big(\Pr(r,\mc{m}{x}) \big)\mathrm{d}r +\!
\int_{\min\{\tilde{R},\theta^{-2/\alpha}\}}^{\theta^{-2/\alpha}}\hspace{-17mm}r
\mathrm{S}\big(\Pr(r,\mc{m}{x}) \big)\mathrm{d}r \!\Bigg)
\\ &\skiph\le\skiph&2\pi\mc{\lambda}{x} \Bigg(\! \int_{0}^{\min\{\tilde{R},
\theta^{-2/\alpha}\}}\hspace{-16mm}r
\mathrm{S}\big(\Pr(r,\mc{\rho}{\ml}) \big)\mathrm{d}r +\!
\int_{\min\{\tilde{R},\theta^{-2/\alpha}\}}^{\theta^{-2/\alpha}}\hspace{-10mm}r/2 \mathrm{d}r \Bigg)\label{eqn:S_ix_1}
\\ &\skiph\le\skiph&
2\pi\mc{\lambda}{x} \int_{0}^{\theta^{-2/\alpha}}
\frac{r e^{\frac{-\pi r^2 \mc{\lambda}{\ml}}{2}}\big(e\pi r^2 \mc{\lambda}{\ml}\big)^{ \frac{\mc{\rho}{\ml}}{2}   }}{ \big(\mc{\rho}{\ml})^{\frac{\mc{\rho}{\ml}}{2}   }}\mathrm{d}r\label{eqn:S_ix_2}
\\&\skiph\le\skiph& \frac{4\mc{\lambda}{x}\sqrt{\pi\mc{\rho}{\ml}}}{\mc{\lambda}{l}}(1+\frac{1}{6\mc{\rho}{\ml}}),
\label{eqn:S_ix_3}
\end{eqnarray}
where \eqref{eqn:S_ix_1} is true because of Lemma~\ref{lem:J1} and the facts that (a) $\mathrm{S}\big(\Pr(r,\mc{m}{x}) \big)$ is a decreasing function of $\mc{m}{x}$, (b) $\mathrm{S}(p)=\sqrt{p(1-p)}$ is a increasing function in $[0, \frac{1}{2}]$ and (c) $\mathrm{S}(p)\le \frac{1}{2}$; \eqref{eqn:S_ix_2} is true because of Lemma~\ref{lem:J2} and $\mc{m}{x}\ge 1$.

From \eqref{eqn:ixv_2} and \eqref{eqn:S_ix_3},
\begin{align}
\mathbb{S}\big\{\mc{i}{x}\big\}\!\le\! \frac{4\mc{\lambda}{x}\!\sqrt{\pi\min\{\mc{m}{x},\mc{\rho}{\ml} \}}}{\mc{\lambda}{l}}(1\!+\!\frac{1}{6\min\{\mc{m}{x},\mc{\rho}{\ml} \}}).\label{eqn:ixv_3}
\end{align}

From \eqref{eqn:Il2}, \eqref{eqn:ix_4}, and \eqref{eqn:ixv_3}, \eqref{eqn:eIl} and \eqref{eqn:vIl} are obtained.

\section{Proof of Theorem~\ref{thm:sDoF}}
\label{pf_thm:sDoF}
From Lemma~\ref{lem:SDI}, to characterize $D^{\mathrm{S}}$,  it is necessary to characterize $\mc{S}{\ml}$ and $\mc{S}{e}$. Since $\mc{S}{\ml}$ is addressed by Lemma~\ref{lem:dim_l}, the focus is on $\mc{S}{e}$.
\begin{Lem}[Characterization of $\mc{S}{e}$] $\mc{S}{e}=\min\big\{\mc{d}{\ml},[\mc{N}{e}-\mc{I}{e}]^+\}$, where $\mc{I}{e} = \mc{d}{\ml}\sum_{\mathbf{b}\in\mc{\mathcal{T}}{\ml}}
\mathbb{I}\{L(\mathbf{e},\mathbf{b})>0\} + \mc{d}{j}\sum_{\mathbf{b}\in\mc{\mathcal{T}}{j}}
\mathbb{I}\{L(\mathbf{e},\mathbf{b})>0\}$, $\mathbb{E}\{\mc{I}{e}\}=\mc{\rho}{\ml}\mc{d}{\ml}+\mc{\rho}{j}\mc{d}{j}$ and  $\mathbb{V}\{\mc{I}{e}\}=\mc{\rho}{\ml}\big(\mc{d}{\ml}\big)^2+\mc{\rho}{j}\big(\mc{d}{j}\big)^2$.
\label{lem:dim_e}
\end{Lem}
\begin{IEEEproof}
The position of LTs and LJs are given by PPPs with density $\mc{\lambda}{\ml}$ and $\mc{\lambda}{j}$, respectively. Hence, $\sum_{\mathbf{b}\in\mc{\mathcal{T}}{\ml}} \mathbb{I}\{L(\mathbf{e},\mathbf{b})>0\}$ and $\mc{d}{j}\sum_{\mathbf{b}\in\mc{\mathcal{T}}{j}} \mathbb{I}\{L(\mathbf{e},\mathbf{b})>0\}$ are independent random variables following Poisson distribution, with parameters $\mc{\rho}{\ml}$ and $\mc{\rho}{j}$, respectively. From the properties of Poisson distribution and \eqref{eqn:Se}, Lemma~\ref{lem:dim_e} is proved.
\end{IEEEproof}

Now start the main flow of the proof of Theorem~\ref{thm:sDoF}.

When $\mc{R}{e}=1-\frac{\mc{N}{e}}{\mc{\rho}{\ml}\mc{d}{\ml}+\mc{\rho}{j}\mc{d}{j}}>
\sqrt{\frac{\max\left\{\mc{d}{\ml},\mc{d}{j}\right\}
\max\left\{\mc{\rho}{\ml},\mc{\rho}{j}\right\}
}{\big(\mc{\rho}{\ml} \big)^{2}\mc{d}{\ml}}}$, from Lemma~\ref{lem:dim_e} and  Chebyshev inequality,
\begin{eqnarray}
\hspace{-10mm}\nonumber\Pr\{\mc{S}{e}\ge 0 \}&\skiph\le\skiph& \frac{\mc{\rho}{\ml}\big(\mc{d}{\ml}\big)^2+\mc{\rho}{j}\big(\mc{d}{j}\big)^2}
{\big(\mc{R}{e}\big)^2\big(\mc{\rho}{\ml}\mc{d}{\ml}+\mc{\rho}{j}\mc{d}{j}\big)^2}
\\&\skiph<\skiph&
\frac{\max\left\{\mc{d}{\ml},\mc{d}{j}\right\}\max\left\{\mc{\rho}{\ml},
\mc{\rho}{j}\right\}}
{\big(\mc{R}{e}\mc{\rho}{\ml}\big)^2\mc{d}{\ml}}.
\label{eqn:Pr_Se1}
\end{eqnarray}

Otherwise, when $\mc{R}{e}< -\sqrt{\frac{\max\left\{\mc{d}{\ml},\mc{d}{j}\right\}
\max\left\{\mc{\rho}{\ml},\mc{\rho}{j}\right\}
}{\big(\mc{\rho}{\ml} \big)^{2}\mc{d}{\ml}}}$, noting that $\mc{\rho}{\ml}\ge 1$,
\begin{eqnarray}
\nonumber\Pr\{\mc{S}{e}\ge \mc{d}{\ml} \}\hspace{-22mm}
\\\nonumber &\skiph\ge&\skiph 1-\frac{\mc{\rho}{\ml}\big(\mc{d}{\ml}\big)^2+\mc{\rho}{j}\big(\mc{d}{j}\big)^2}
{\big(\mc{R}{e}\big(\mc{\rho}{\ml}\mc{d}{\ml}+\mc{\rho}{j}\mc{d}{j}\big)+\mc{d}{\ml}\big)^2}
\\\nonumber &\skiph\ge&\skiph 1-\frac{\mc{\rho}{\ml}\big(\mc{d}{\ml}\big)^2+\mc{\rho}{j}\big(\mc{d}{j}\big)^2}
{\big(\mc{R}{e}\big(\mc{\rho}{\ml}\mc{d}{\ml}+\mc{\rho}{j}\mc{d}{j}\big)\big)^2}+
\\\nonumber&&\hspace{3mm}
\frac{2\left(\mc{\rho}{\ml}\big(\mc{d}{\ml}\big)^2+\mc{\rho}{j}\big(\mc{d}{j}\big)^2\right)\mc{d}{\ml}}
{\big(\mc{R}{e}\big(\mc{\rho}{\ml}\mc{d}{\ml}+\mc{\rho}{j}\mc{d}{j}\big)\big)^3
}
\\&\skiph\sim&\skiph 1-  \mathcal{O}\left(\frac{\max\left\{\mc{d}{\ml},\mc{d}{j}\right\}\max\left\{\mc{\rho}{\ml},
\mc{\rho}{j}\right\}}
{\big(\mc{R}{e}\mc{\rho}{\ml}\big)^2\mc{d}{\ml}}\right).\label{eqn:Pr_Se2}
\end{eqnarray}

Similarly, from Lemma~\ref{lem:dim_l} and  Chebyshev inequality, when $\mc{R}{\ml}=\frac{\mc{N}{\ml}\!-\!\mc{d}{\ml} +
\min\{\mc{m}{\ml}\mc{d}{\ml},\mc{\rho}{\ml}\mc{d}{\ml}\} +\min\{\frac{\mc{\lambda}{j}}{\mc{\lambda}{l}}\mc{m}{j}\mc{d}{j},
\mc{\rho}{j}\mc{d}{j}\}}{\mc{\rho}{\ml}\mc{d}{\ml}+\mc{\rho}{j}\mc{d}{j}}
-1>\sqrt{\frac{\max\left\{\mc{d}{\ml},\mc{d}{j}\right\}
\max\left\{\mc{\rho}{\ml},\mc{\rho}{j}\right\}
}{\big(\mc{\rho}{\ml} \big)^{2}\mc{d}{\ml}}}$,
\begin{eqnarray}
\hspace{-3mm}\Pr\{\mc{S}{\ml}< \mc{d}{\ml} \}\sim \mathcal{O}\bigg(\frac{\max\left\{\mc{d}{\ml},\mc{d}{j}\right\}\max\left\{\mc{\rho}{\ml},
\mc{\rho}{j}\right\}}
{\big(\mc{R}{e}\mc{\rho}{\ml}\big)^2\mc{d}{\ml}}\bigg), \label{eqn:Pr_Sl1}
\end{eqnarray}
and when $\mc{R}{\ml}<-\sqrt{\frac{\max\left\{\mc{d}{\ml},\mc{d}{j}\right\}
\max\left\{\mc{\rho}{\ml},\mc{\rho}{j}\right\}
}{\big(\mc{\rho}{\ml} \big)^{2}\mc{d}{\ml}}}$,
\begin{eqnarray}
\Pr\{\mc{S}{\ml}\ge 0 \}\sim 1\!-\!\mathcal{O}\bigg(\!\frac{\max\left\{\mc{d}{\ml}\!,\!\mc{d}{j}
\right\}\max\left\{\mc{\rho}{\ml}\!,\!
\mc{\rho}{j}\right\}}
{\big(\mc{R}{e}\mc{\rho}{\ml}\big)^2\mc{d}{\ml}}\!\bigg). \label{eqn:Pr_Sl2}
\end{eqnarray}

Substituting \eqref{eqn:Pr_Se1}--\eqref{eqn:Pr_Sl2} to \eqref{eqn:sDoF2}, \eqref{eqn:sdof_T} is obtained.

\section{Feasible Region under High Connection Density}
\label{sec:dense}

When $\mc{\rho}{\ml}\gg 1$, the feasible streams $(\mc{d}{j},\mc{d}{\ml})\in\left\{0,\ldots,\mc{M}{j}\right\}\times\left\{1,\ldots,\min\left\{\mc{N}{\ml},\mc{M}{\ml}\right\}\right\}
$ need to satisfy
\begin{eqnarray}
R\hspace{3.4mm}&\skiph=\skiph&\min\{\mc{R}{e},\mc{R}{\ml}\}> 0,\qquad\mbox{where}\label{eqn:Rfeasible}\\
\mc{R}{e}&\skiph=\skiph&1-\frac{\mc{N}{e}}{\mc{\rho}{\ml}\mc{d}{\ml}+\mc{\rho}{j}\mc{d}{j}},
\label{eqn:Re}\\
\mc{R}{\ml}&\skiph=\skiph&\frac{\mc{N}{\ml}\!-\!\mc{d}{\ml} +
\min\{\mc{m}{\ml}\mc{d}{\ml},\mc{\rho}{\ml}\mc{d}{\ml}\}}{\mc{\rho}{\ml}\mc{d}{\ml}+
\mc{\rho}{j}\mc{d}{j}}\nonumber
\\&&\frac{+\min\{\frac{\mc{\rho}{j}}{\mc{\rho}{l}}\mc{m}{j}\mc{d}{j},
\mc{\rho}{j}\mc{d}{j}\}}{}-1.\label{eqn:Rl}
\end{eqnarray}

Define
\begin{eqnarray}
\mc{\tilde{R}}{\ml}=\frac{\mc{N}{\ml} +
\min\{\mc{M}{\ml},\mc{\rho}{\ml}\mc{d}{\ml}\}+
}{\mc{\rho}{\ml}\mc{d}{\ml}+\mc{\rho}{j}\mc{d}{j}}\hspace{10mm}\nonumber
\\\frac{\min\{\frac{\mc{\rho}{j}(\mc{M}{j}\!-\!\mc{d}{j})}{\mc{\rho}{\ml}\mc{d}{\ml}}\mc{d}{j},
\mc{\rho}{j}\mc{d}{j}\}}{}-1.\label{eqn:tRl}
\end{eqnarray}
Since the quantization error of the $\lfloor\cdot\rfloor$ function is bounded by $(-1,0]$,
by substituting \eqref{eqn:ml} to \eqref{eqn:Rl}, we get
\begin{eqnarray}
\left|\mc{R}{\ml}-\mc{\tilde{R}}{\ml}\right|\le
\frac{2\mc{d}{\ml}+
\frac{\mc{\rho}{j}}{\mc{\rho}{\ml}}\mc{d}{j}}{\mc{\rho}{\ml}\mc{d}{\ml}+
\mc{\rho}{j}\mc{d}{j}}\le\frac{2}{\mc{\rho}{\ml}}.
\end{eqnarray}
Hence, the difference between $\mc{R}{\ml}$ and $\mc{\tilde{R}}{\ml}$ is ignorable when $\mc{\rho}{\ml}\gg 1$.
Therefore, one can replace $\mc{R}{\ml}$ by $\mc{\tilde{R}}{\ml}$ in \eqref{eqn:Rfeasible}.
After this replacement, \eqref{eqn:Rfeasible} is equivalent to the following four inequalities:
\begin{eqnarray}
\hspace{-1mm}f_1(\mc{d}{j},\mc{d}{\ml})&\skiph=&\skiph
\mc{\rho}{\ml}\mc{d}{\ml}+\mc{\rho}{j}\mc{d}{j}-\mc{N}{e}> 0,\label{eqn:jam}\\
\hspace{-1mm}f_2(\mc{d}{j},\mc{d}{\ml})&\skiph=&\skiph
\mc{\rho}{\ml}\mc{d}{\ml}-\mc{N}{\ml}-\mc{M}{\ml}< 0,\label{eqn:clear1}\\
\hspace{-1mm}f_3(\mc{d}{j},\mc{d}{\ml})&\skiph=
&\skiph\big(\mc{\rho}{\ml}\big)^2
\big(\mc{d}{\ml}\big)^2
+\mc{\rho}{\ml}\mc{\rho}{j}\mc{d}{\ml}\mc{d}{j}
+\mc{\rho}{j}\big(\mc{d}{j}\big)^2\hspace{-2mm}-\nonumber\\
&&\skiph(\mc{M}{\ml}+\mc{N}{\ml})\mc{\rho}{\ml}\mc{d}{\ml}
-\mc{M}{j}\mc{\rho}{j}\mc{d}{j}<0,\label{eqn:clear2}\\
\hspace{-1mm}f_4(\mc{d}{j},\mc{d}{\ml})&\skiph=&\skiph
\mc{\rho}{\ml}\mc{\rho}{j}\mc{d}{\ml}\mc{d}{j}
+\mc{\rho}{j}\big(\mc{d}{j}\big)^2
-\mc{N}{\ml}\mc{\rho}{\ml}\mc{d}{\ml}\nonumber\\
&&\skiph-\mc{M}{j}\mc{\rho}{j}\mc{d}{j}<0.\label{eqn:clear3}
\end{eqnarray}

It is easy to see that $f_1(\mc{d}{j},\mc{d}{\ml})=0$ is a line with slope $-\frac{\mc{\rho}{j}}{\mc{\rho}{\ml}}$ and intersection $\frac{\mc{N}{e}}{\mc{\rho}{\ml}}$, and $f_2(\mc{d}{j},\mc{d}{\ml})=0$ is a horizontal line with intersection $\frac{\mc{M}{\ml}+\mc{N}{\ml}}{\mc{\rho}{\ml}}$.
However, as $f_3(\mc{d}{j},\mc{d}{\ml})=0$ and $f_4(\mc{d}{j},\mc{d}{\ml})=0$ are second order curves, it is difficult to give a simple characterization of the feasible region.
On the other hand, noting that
\begin{itemize}
\item[--] $f_3(\mc{d}{j},\mc{d}{\ml})=0$ passes through points
$(0,0)$, $(0,\frac{\mc{M}{\ml}+\mc{N}{\ml}}{\mc{\rho}{\ml}})$, $(\max\{\mc{M}{j}-\mc{M}{\ml}-\mc{N}{\ml},0\},$ $\frac{\mc{M}{\ml}+\mc{N}{\ml}}{\mc{\rho}{\ml}})$ and $(\mc{M}{j},0)$,
\item[--] $f_4(\mc{d}{j},\mc{d}{\ml})=0$ passes through points $(0,0)$, $(\mc{M}{j},0)$,
\end{itemize}
we have the following proposition to summarize the property of $f_2(\mc{d}{j},\mc{d}{\ml})$, $f_3(\mc{d}{j},\mc{d}{\ml})$ and $f_4(\mc{d}{j},\mc{d}{\ml})$.
\begin{Prop}[Properties of the Feasible Region] When $\mc{M}{j}\ge\mc{M}{\ml}+\mc{N}{\ml}$,
the points $(\mc{d}{j},\mc{d}{\ml})$ in the interior of trapezoid with vertices $(0,0)$, $(0,\frac{\mc{M}{\ml}+\mc{N}{\ml}}{\mc{\rho}{\ml}})$, $(\mc{M}{j}-\mc{M}{\ml}-\mc{N}{\ml},\frac{\mc{M}{\ml}+\mc{N}{\ml}}{\mc{\rho}{\ml}})$ and $(\mc{M}{j},0)$ satisfy \eqref{eqn:clear1}--\eqref{eqn:clear3}.
\end{Prop}
\begin{IEEEproof}
Noting that function $f(x)=\frac{a+x}{b+x}$, with $a,b\ge0$ and $x\in[0,+\infty)$ is a strictly decreasing function when $f(x)> 1$ (i.e., $a> b$), from \eqref{eqn:tRl}, $\mc{\tilde{R}}{\ml}$ is an strictly decreasing function of $\mc{d}{j}$ and $\mc{d}{\ml}$ when $\mc{\tilde{R}}{\ml}> 0$.
Further noting that $\eqref{eqn:clear1}$--$\eqref{eqn:clear3} \Leftrightarrow \mc{\tilde{R}}{\ml}> 0$, if $\eqref{eqn:clear1}$--$\eqref{eqn:clear3}$ hold for certain data stream configuration $(\mc{d}{j}, \mc{d}{\ml})$, then for any $(\mc{\tilde d}{j}, \mc{\tilde d}{\ml})$ satisfying $\mc{\tilde d}{j}\le \mc{d}{j}$ and $\mc{\tilde d}{\ml}\le \mc{d}{\ml}$, $\eqref{eqn:clear1}$--$\eqref{eqn:clear3}$ must hold.
Therefore, to prove the proposition, one only need to prove that
$f_i(\mc{d}{j},\mc{d}{\ml})\le0$, $i\in\{2,3,4\}$ for all points on the line segment with end points $(\mc{M}{j}-\mc{M}{\ml}-\mc{N}{\ml},\frac{\mc{M}{\ml}+\mc{N}{\ml}}{\mc{\rho}{\ml}})$ and $(\mc{M}{j},0)$. Since on this line segment, $\mc{d}{\ml}\le \frac{\mc{M}{\ml}+\mc{N}{\ml}}{\mc{\rho}{\ml}}$, it is clear that $f_2(\mc{d}{j},\mc{d}{\ml})\le0$. This line segment can be expressed as
\begin{eqnarray}
\mc{d}{\ml}=\frac{\mc{M}{j}-\mc{d}{j}}{\mc{\rho}{\ml}},\ \mc{d}{j}\in\{\mc{M}{j}-\mc{M}{\ml}-\mc{N}{\ml},\mc{M}{j}\}.\label{eqn:lineseg}
\end{eqnarray}
Substitute \eqref{eqn:lineseg} into \eqref{eqn:clear2} and  \eqref{eqn:clear3},
\begin{align}
\nonumber f_3(\mc{d}{j},\mc{d}{\ml})=&(\mc{d}{j} -\mc{M}{j})
(\mc{d}{j} +\mc{N}{\ml} +\mc{M}{\ml} -\mc{M}{j}),\\
\nonumber f_4(\mc{d}{j},\mc{d}{\ml})=&\mc{N}{\ml}(\mc{d}{j} -\mc{M}{j}).
\end{align}
Hence, when $\mc{d}{j}\in\{\mc{M}{j}-\mc{M}{\ml}-\mc{N}{\ml},\mc{M}{j}\}$, $f_i(\mc{d}{j},\mc{d}{\ml})\le0$, $i\in\{3,4\}$. This completes the proof.
\end{IEEEproof}


\end{document}